\documentclass[pdftex,numberedappendix,appendixfloats,apj]{emulateapj}

\begin{document}

\shorttitle{Metal diffusion in SPH dwarf galaxies}

\shortauthors{Williamson, Martel \& Kawata}

\title{Metal diffusion in smoothed particle hydrodynamics simulations of dwarf galaxies}

\author{David Williamson}
\affil{D\'epartement de physique, de g\'enie physique et d'optique, Universit\'e Laval, Qu\'ebec, QC, G1V 0A6, Canada}
\affil{Centre de Recherche en Astrophysique du Qu\'ebec, QC, Canada}
\email{david-john.williamson.1@ulaval.ca}

\author{Hugo Martel}
\affil{D\'epartement de physique, de g\'enie physique et d'optique, Universit\'e Laval, Qu\'ebec, QC, G1V 0A6, Canada}
\affil{Centre de Recherche en Astrophysique du Qu\'ebec, QC, Canada}


\author{Daisuke Kawata}
\affil{Mullard Space Science Laboratory, University College London, Holmbury St Mary, Dorking, Surrey, UK}

\begin{abstract}
We perform a series of smoothed particle hydrodynamics simulations of isolated dwarf galaxies to compare different metal mixing models. In particular, we examine the role of diffusion in the production of enriched outflows, and in determining the metallicity distributions of gas and stars. We investigate different diffusion strengths, by changing the pre-factor of the diffusion coefficient, by varying how the diffusion coefficient is calculated from the local velocity distribution, and by varying whether the speed of sound is included as a velocity term. Stronger diffusion produces a tighter [O/Fe]-[Fe/H] distribution in the gas, and cuts off the gas metallicity distribution function at lower metallicities. Diffusion suppresses the formation of low-metallicity stars, even with weak diffusion, and also strips metals from enriched outflows. This produces a remarkably tight correlation between ``metal mass-loading'' (mean metal outflow rate divided by mean metal production rate) and the strength of diffusion, even when the diffusion coefficient is calculated in different ways. The effectiveness of outflows at removing metals from dwarf galaxies and the metal distribution of the gas is thus dependent on the strength of diffusion. By contrast, we show that the metallicities of stars are not strongly dependent on the strength of diffusion, provided that some diffusion is present.
\end{abstract}


\section{Introduction}

It has been well-established that metals are not spread homogeneously in galaxies and the surrounding medium. Vertical and radial metallicity gradients have been observed in galaxies \citep{1983MNRAS.204...53S,1994ApJ...420...87Z,2010ApJS..190..233M,2012ApJ...745...66M}, as well as bimodality in the metallicity of the circumgalactic medium (CGM) \citep{2013ApJ...770..138L}. However, the details of where metals are produced and distributed across a galaxy remain unclear. Observations suggest that the radial metallicity gradient can steepen, flatten, or even reverse over time \citep{2003A&A...397..667M,2008ApJ...674..172R,2010ApJ...714.1096S}. Feedback-driven outflows are a major contributor to this behaviour, by enriching the CGM and removing metals from galaxies, they likely play a key role in establishing the Mass-Metallicity relation \citep{2004ApJ...613..898T,2008MNRAS.387..577O}.

These metal distributions can be a powerful constraint for hydrodynamic simulations, potentially alleviating the degeneracy between simulations that produce similar hydrodynamic and kinematic results. For these constraints to be useful, we must examine the production and distribution of metals in simulated galaxies, and perform an extensive analysis of any relevant numerical issues. This is a broad subject, and so in this paper, we focus primarily on the effects of diffusion on the metal distribution in smoothed particle hydrodynamics (SPH) simulations of dwarf galaxies with strong outflows.

There exists extensive research into analytic or one-zone chemical evolution models \citep[e.g.][]{1986A&A...154..279M,1995A&A...304...11M,1995ApJS...98..617T,2004A&A...421..613F,2004MNRAS.347..968P,2005A&A...430..491R,2006MNRAS.365.1114P,2007A&A...468..927L}, but semi-analytic chemical evolution models \citep[e.g.][]{2005ApJ...634...26N,2009A&A...505.1075P,2010MNRAS.402..173A,2013ApJ...777..107C,2015arXiv150402109L} and {\em chemodynamic} models \citep[e.g.][]{2003MNRAS.340..908K,2005MNRAS.364..552S,2007MNRAS.376.1465K,2008MNRAS.390.1349P,2009ApJ...697...55G} have been developed more recently. By including three-dimensional hydrodynamics and resolved chemical enrichment and mixing, chemodynamic models do not require the strong assumptions regarding metal production and wind properties that are necessary for one-zone or semi-analytic models. Furthermore, these allow us to examine the spatial distribution of metals, and to relax the instantaneous recycling approximation.

The distribution of metals in simulations depends strongly on sub-grid models of feedback, metal injection, and metal mixing, which can differ greatly between different simulation codes. Cosmological simulations often include prescriptions for the generation of wind particles \citep[e.g.][]{2008MNRAS.387..577O}, because resolution limits prevent the explicit formation of winds from stellar feedback. Simulations of isolated galaxies, and higher resolution cosmological simulations can explicitly resolve some of these processes, somewhat reducing the dependence of the results on the details of sub-grid models, but a dependence still remains even at high resolutions.

The effects of these sub-grid processes are tightly coupled with each other. The negative gas surface-density gradient of a galaxy favors star formation (and hence metal injection) that is centrally concentrated -- the ``inside-out'' galaxy formation scenario \citep{2012A&A...540A..56P} -- producing a steep negative metallicity gradient in the absence of strong metal mixing. At the same time, stellar feedback pushes this metal-rich gas outwards, with stronger feedback resulting in shallower metallicity gradients \citep{2012A&A...540A..56P,2013A&A...554A..47G}. Feedback can temporarily eject the gas from dwarf galaxies, resulting in episodic star formation \citep{2007ApJ...667..170S,2014MNRAS.444.3845F}. Strong feedback can produce an irregular density distribution, spreading out star formation across the disk \citep[][hereafter K14]{2014MNRAS.438.1208K}, with a strong effect on metal distribution. Rapid metal diffusion can remove metals from rich outflows, reducing the effectiveness of metal advection from feedback \citep{2004MNRAS.352..363M}, where we define ``advection'' as the transport of metals through the movement of metal-enriched flows of gas, as opposed to diffusion, which can transport metals without any movement of gas. The flow of metals and rate of metal injection sets the metallicity slope of the disk, which affects the local cooling time and hence feeds back into the star formation and from there into the metal injection rate. Thus it is important to examine these processes in hydrodynamic simulations to determine the details of these interactions. In particular, the effectiveness of this metal flow in simulations depends on whether a diffusion algorithm is included, and if present, the choice of the diffusion algorithm and its calibration. This is especially important for SPH simulations where there is no intrinsic diffusion of metals, and it is within this context that we examine diffusion in detail in this work.

The rest of this paper is organized as follows. In section~\ref{section_method}, we describe our simulation code, initial conditions, and the diffusion models we are investigating, and provide a discussion on star formation and feedback. In section~\ref{section_results} we present the results of these simulations, and examine the effects that the choice of diffusion has on the evolution of the metallicity distributions. In section~\ref{hydrodiscuss} we perform a brief analysis of unresolved processes. In section~\ref{section_conc} we summarize our conclusions. In appendix~\ref{converge}, we discuss the results of a higher-resolution model as a convergence check, and in appendix~\ref{diffcheck} we make use of idealized particle distributions to investigate the fundamental differences between diffusion models.

\section{Method}\label{section_method}

\subsection{Simulation Code}

The choice of hydrodynamics scheme can have a significant effect on galaxy evolution. Eulerian hydrodynamics codes implicitly include some numerical diffusion, which can produce mixing that is implementation-dependent, and often strong \citep{1999MNRAS.309..941D,2001MNRAS.322..800R,2002ApJ...581.1047D}. Lagrangian methods such as SPH require an explicit mixing scheme, and so diffusion is completely dependent on the choice of diffusion model. Standard SPH methods also typically suppress the Rayleigh-Taylor (RT) and Kelvin-Helmholtz (KH) instabilities \citep{2007MNRAS.380..963A}. These instabilities can contribute to the production of turbulence and gas mixing, and their absence could reduce the effectiveness of feedback. However, the lack of intrinsic mixing in SPH permits complete control over the level of metal diffusion, even making it possible to completely switch off sub-grid diffusion. We use an SPH code in this work to fully investigate the effects of metal diffusion in our simulations. Adjustments to SPH have also improved its ability to capture the RT and KH instabilities \citep{2008JCoPh.22710040P}, and our simulation code makes use of these improvements \citep{2013MNRAS.428.1968K}.

We perform simulations of isolated dwarf galaxies to better examine the mechanisms of metal distribution and wind production without the complexity of a cosmological context. Our simulations are performed with the numerical simulation algorithm \textsc{GCD+} \citep[][K14]{2003MNRAS.340..908K,2012MNRAS.420.3195B,2013MNRAS.428.1968K}, an MPI TreeSPH code that includes self-consistent stellar feedback, radiative cooling, a dynamic temperature floor, artificial thermal conductivity, and explicit metal production and diffusion. Details of the latest version of this code can be found in K14. Here we summarize some of the salient features of the code, and the modifications made for this study.

\subsection{Feedback, star formation, and cooling}

Our feedback algorithm is only slightly modified from that described in detail in K14, and includes a UV background and stellar feedback. Briefly, gas particles are converted into star particles according to a stochastic method based on the local star formation time-scale. The stellar initial mass function (IMF) is divided into $61$ mass groups, and each star particle is assigned to one of these groups. These star particles then inject energy and metals through feedback, according to their mass group. This feedback represents stellar winds, Type-II supernovae, and Type-Ia supernovae. A ``feedback particle'' receives thermal energy according to its age and mass group, and for Type-II supernovae particles, radiative cooling is temporarily turned off. Neighboring particles receive kinetic energy only through the increased pressure of this feedback particle. Switching off cooling in this fashion can perhaps be thought of as a very simple sub-grid turbulence model, as it represents a source of sub-grid internal energy that can not be immediately radiated away, and that eventually cascades into thermal energy. We set the star formation efficiency to $\epsilon_*=0.02$, and use a threshold density of $n_\mathrm{H}^{\phantom i}=1$ cm$^{-3}$. We performed some tests with different values, and found that the star formation rate does not depend strongly on modest changes to these parameters.

\begin{figure}
\begin{center}
\includegraphics[width=\columnwidth]{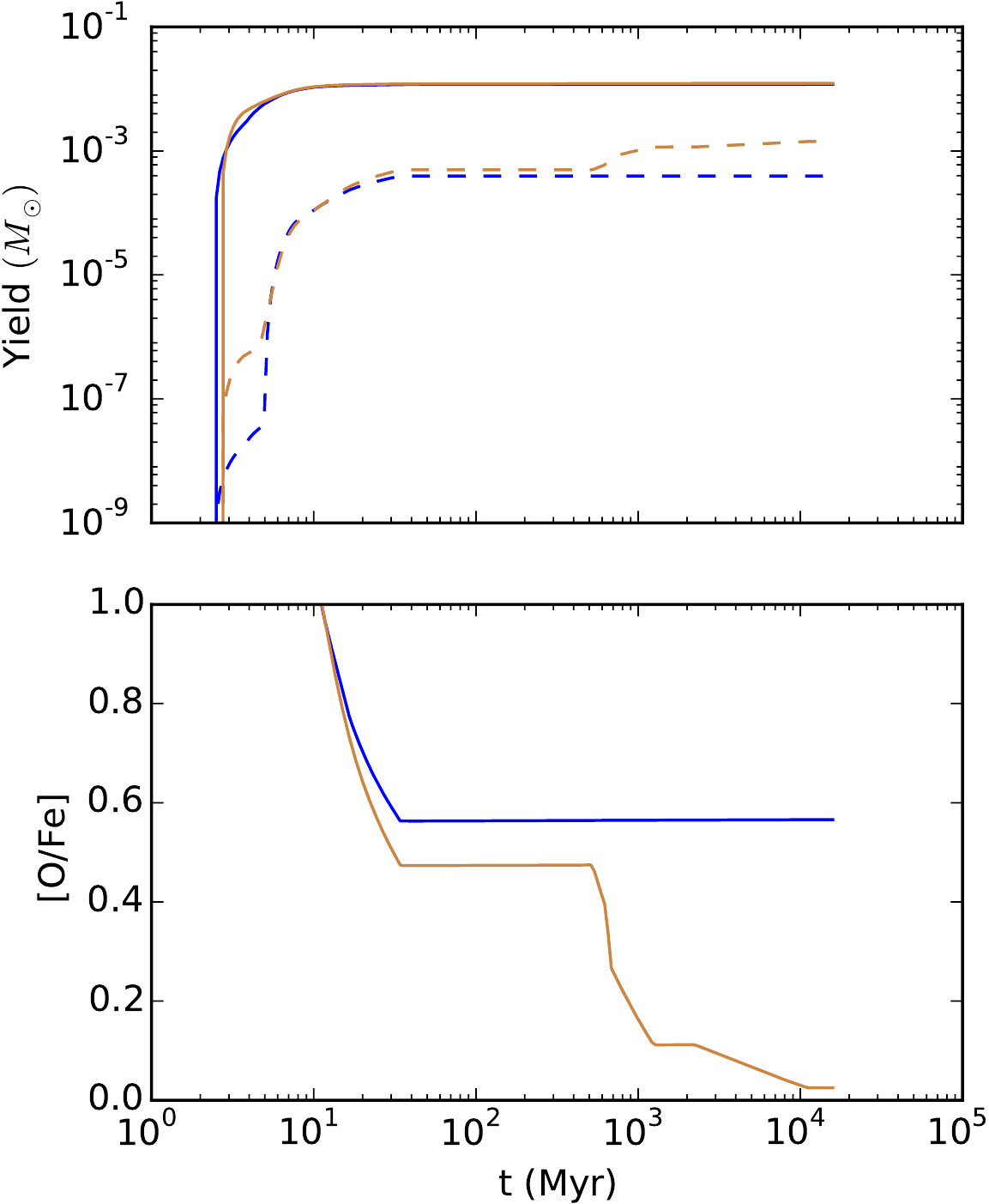}\\
\end{center}
\caption{\label{yieldplot}Top: The IMF-weighted total yields of of oxygen (solid lines) and iron (dashed lines) as a function of age for a star cluster with a total initial mass of $1~M_\odot$ and an initial metallicity of $Z=10^{-2}~Z_\odot$ (blue lines) and $Z=0.11~Z_\odot$ (brown lines). Bottom: IMF-weighted mean [O/Fe] of metals produced with an initial metallicity of $Z=10^{-2~}Z_\odot$ (blue line) and $Z=0.11~Z_\odot$ (brown line).
}
\end{figure}

Metals are produced by star particles according to their mass group. The IMF-weighted total yields of iron and oxygen produced by star particles as a function of age for a star cluster with a total initial mass of $1~M_\odot$, are plotted in Fig.~\ref{yieldplot}. SNe II produce large amounts of iron and oxygen within $10$ Myr, while SNe Ia primarily contribute iron, $\sim1$ Gyr after the star formation event. Following the SNe Ia model of \citet{2000MmSAI..71..461K} who suggest a threshold metallicity for the production of SNe Ia, we only allow the production of SNe Ia from gas particles with $Z/Z_\odot>-1.1$. Therefore, there is no contribution from SNe Ia in the $Z=10^{-2} Z_\odot$ case in Fig.~\ref{yieldplot}.

In the previous version of GCD+ (K14), metals are only injected into the feedback particle, from where they can only propagate through diffusion. We have modified GCD+ so that these metals are injected into surrounding gas following the feedback particle's smoothing kernel. Distributing metals in this fashion helps to promote metal-rich winds, as metals and kinetic energy are added to the same particles (although kinetic energy is only added indirectly through the increased pressure of the feedback particle). 

Radiative heating (i.e. the UV background) and cooling rates are unmodified from those described in K14. These have been tabulated with CLOUDY \citep{1998PASP..110..761F}, and are a function of temperature, metallicity, density, and redshift, although in this work we assume a constant redshift of $z=0$. As in K14, a pressure floor is also included to avoid numerical instabilities \citep{1997MNRAS.288.1060B}. The thermal energy is updated by solving the entropy equation with an implicit method, described in \citet{2006ApJ...641..785K}, circumventing the need for short time-steps when cooling is rapid.

\subsection{Metal diffusion}

Our metal diffusion algorithm is based on the method of \citet{2009MNRAS.392.1381G}, which uses an implicit scheme to produce accurate diffusion even at large time-steps. We modify this algorithm by using different methods for calculating the diffusion coefficient of a particle. For our newly implemented ``Shear'' method, we follow the method of \citet[][hereafter S10]{2010MNRAS.407.1581S} based on \citet{1963MWRv...91...99S} for calculating the diffusion coefficient. We stress that we have not implemented the general diffusion algorithm of S10, but merely implement their method for calculating the diffusion coefficient, within the diffusion scheme of \citet{2009MNRAS.392.1381G}. Here, the diffusion coefficient of particle $i$ is

\begin{equation}
D_{i,S} = C_{DS}|S_{ab,i}|h_i^2,
\end{equation}
where $h_i$ is the smoothing length of particle $i$, the pre-factor $C_{DS}$ is a scaling constant for this ``shear'' diffusion, and $S_{ab,i}$ is the trace-free symmetric shear tensor for particle $i$. The shear tensor $S_{ab,i}$ is calculated from the kernel-weighted sum over all neighbors $j$,

\begin{eqnarray}
\tilde{S}_{ab,i} &=& \frac{1}{\rho_i}\sum_j m_j (v_{b,i}^{\phantom i} - v_{b,j}^{\phantom i}) \nabla_{i,a} W_{ij},\\
S_{ab,i} &=& \frac{1}{2} (\tilde{S}_{ab,i} + \tilde{S}_{ba.i}) - \delta_{ab}\frac{1}{3} \mathrm{Tr} (\mathbf{\tilde{S}}),
\end{eqnarray}
(S10), where $\nabla_{i,a}W_{ij}$ is $a$th component of the gradient of the SPH kernel function for particle $i$ (i.e. $a=x,y,$ or $z$), $v_{b,i}^{\phantom i}$ and $v_{b,j}^{\phantom i}$ are the $b$th component of the physical velocity vector of particles $i$ and $j$ (where again, $b=x,y,$ or $z$), $\delta_{ab}$ is the Kronecker delta, $\rho_i$ is the density of particle $i$, and $m_j$ is the mass of particle $j$. We set $C_{DS}$=0.1. S10 states that $C_{DS}=0.05-0.1$ is expected from turbulence theory, and selects a value of $C_{DS}=0.05$, and so our diffusion is somewhat stronger. The S10 method is designed to prevent diffusion from occurring in situations such as solid body rotation, or in a purely laminar expansion or compression -- situations that generate significant diffusion in a method based solely on velocity dispersion. We discuss the effectiveness of this in appendix~\ref{diffcheck}.

Previous versions of \textsc{GCD+} calculated the diffusion coefficient using the simpler ``velocity-dispersion'' model of \citet{2009MNRAS.392.1381G}. It has been noted that this metal diffusion was perhaps too strong and results in too small a dispersion in stellar metallicities (K14). To quantify this better, we compare this model with the Shear method above. We also experiment with including a sound-speed term in this model. For brevity, we use ``Dispnoc'' to refer to the velocity-dispersion model that excludes the sound-speed term, and ``Disp'' to refer to the velocity-dispersion model and includes the sound-speed term. In Disp and Dispnoc, the diffusion coefficient of a particle is given as:

\begin{equation}
D_{i,V} = C_{DV} V_i h_i,
\end{equation}
where the pre-factor $C_{DV}$ is a scaling constant for this diffusion model, and $V_i$ is a velocity scale for particle $i$. For the Dispnoc model, $V_i$ is defined to equal to the local velocity dispersion, $v_{i,\mathrm{disp}}^{\phantom i}$, calculated by a sum of the velocity difference over all neighboring particles $j$,
\begin{equation}
v_{i,\mathrm{disp}}^2 = \frac{1}{N}\sum_j|\mathbf{v_i^{\phantom i}}-\mathbf{v_j^{\phantom i}}|^2,
\end{equation}
where $\mathbf{v_i^{\phantom i}}$ and $\mathbf{v_j^{\phantom i}}$ are the physical velocities of particles $i$ and $j$.

For the Disp model, we define $V_i$ to be equal to the quadrature sum of the velocity dispersion and the sound speed, i.e. $V_i = (c_i^2+v_{i,\mathrm{disp}}^2)^{1/2}$. In practice, we found this has a negligible effect on most of the disk of a simulated galaxy, except in gas particles that neighbor feedback particles. For the very hot feedback particles, the diffusion coefficient is significantly increased, which could be thought of as representing a rapid mixing of metals due to unresolved turbulence produced by stellar winds and supernovae.

We vary $C_{DV}$, setting $C_{DV}=1$ as in K14, or $C_{DV}=0.1$ for better comparison with the Shear model. \citet{2009MNRAS.392.1381G} set $C_{DV}=2$, deriving this from the theoretical calculations of \citet{2003PhRvE..67d6311K}, but in practice such a large value results in extremely rapid diffusion.

Combining these, we have five different methods for calculating the diffusion coefficient: Shear, Disp with $C_{DV}=1$, Dispnoc with $C_{DV}=1$, Disp with $C_{DV}=0.1$ (which we call ``Displow''), and Dispnoc with $C_{DV}=0.1$ (which we call ``Displownoc'').

\subsection{Simulations}

Simulated galaxies that evolve from idealized axisymmetric disks in hydrodynamic equilibrium (where gravity is perfectly balanced by pressure and orbital motion) can collapse monolithically, uninhibited by feedback because the threshold densities for star formation have not yet been reached. The gas can reach large densities simultaneously throughout the disk, producing a powerful burst of star formation that can disrupt the disk entirely, or generate extremely powerful outflows. Observed galaxies exist in a more dynamical equilibrium, with gas existing in multiple phases within non-axisymmetric features such as bars, spiral arms, clouds, and bubbles, as well as inflow and outflow, and thus real galaxies do not collapse in such a uniform axisymmetric fashion.

Extracting initial conditions (ICs) from a cosmological simulation is a common method that produces galaxies that are stable against such a collapse, and these have the additional advantages of including the detail of the local environments, and being more likely to represent typical examples of galaxies from a desired sample. However, cosmological ICs introduce a complexity and an additional level of model dependency that can make it more difficult to directly compare the roles of the different processes within a galaxy. Galaxy models with cosmological ICs may more effectively capture the large-scale and long-term features of galaxy evolution, but more idealized models can act as a better laboratory to perform specific experiments to understand the details of galaxy evolution and numerical models.

In this work, we produce idealized axisymmetric ICs, and to avoid the problem of an initial burst of star formation, we damp star formation for the first $200$ Myr of evolution, by increasing the star formation time-scale through the relation
\begin{equation}
\tau_\mathrm{SF,damped}^{\phantom X}=\tau_\mathrm{SF}^{\phantom X} (200 \mathrm{Myr}/t),
\end{equation}where $t$ is the time, and $\tau_\mathrm{SF,damped}^{\phantom i}$ and $\tau_\mathrm{SF}^{\phantom i}$ are the damped and undamped star formation time-scales.

Our model does not include the infall of pristine gas, or interactions with other galaxies. Most observed dwarf galaxies appear in groups \citep{1987ApJ...321..280T} where interactions are common, and so these simulations are not intended to represent an average dwarf galaxy at some redshift, but primarily as a laboratory to examine the details of secular chemodynamical evolution in the absence of external disturbances. This means that our results are more universally applicable, as they do not depend on the details of a particular interaction history.

\begin{table}
\begin{tabular}{ccccccc}
\hline\hline
Name & Method & $C_{D}$ & Sound & $N_\mathrm{b}$ & $N_\mathrm{DM}$\\
\hline
Nodiff & None & N/A & N/A & $5\times10^5$ & $9.5\times10^5$ \\
Shear & Shear & 0.1 & No & $5\times10^5$ & $9.5\times10^5$ \\
Displownoc & Disp & 0.1 & No & $5\times10^5$ & $9.5\times10^5$ \\
Displow & Disp & 0.1 & Yes & $5\times10^5$ & $9.5\times10^5$ \\
Dispnoc  & Disp & 1 & No & $5\times10^5$ & $9.5\times10^5$ \\
Disp & Disp & 1 & Yes & $5\times10^5$ & $9.5\times10^5$ \\
Hires & Shear & 0.1 & No & $15\times10^5$ & $28.5\times10^5$ \\
\hline
\end{tabular}
\caption{\label{ictable} \textup{
Summary of simulation parameters. ``Shear'' indicates that the simulation uses the diffusion coefficient based on S10, ``Disp'' indicates that the simulation uses the diffusion coefficient based on \citet{2009MNRAS.392.1381G}, and ``None'' indicates that the simulation does not include metal diffusion. $C_{D}$ is the scaling factor for the diffusion coefficient (i.e. $C_{DV}$ or $C_{DS}$). ``Sound'' indicates whether the model includes the speed of sound in the velocity term of the diffusion coefficient. $N_\mathrm{b}$ and $N_\mathrm{DM}$ are the number of baryonic and dark matter particles in the simulation respectively.
}
}
\end{table}

We performed six different simulations at our fiducial resolution, each with different methods for calculating the diffusion coefficients. The initial conditions have a large gas fraction ($f_g=0.5$) and low initial metallicity ($Z_\mathrm{W}=10^{-2} Z_{\mathrm{W},\odot}$ for all metal species W). The gas fraction of the model is intentionally greater than that of local dwarfs \citep{2009ApJ...696..385G}. This allows the galaxy to form stars, and reach an equilibrium that is more directly a result of the modelled evolution, and less strongly controlled by the precise initial conditions.

The initial metal ratios were set to the solar values, giving a low initial [O/Fe]. To correct this, we apply a post-processing operation where we subtract a fixed mass of iron from every particle (increasing the hydrogen mass by the same amount), such that the initial [O/Fe] of each particle is [O/Fe]$=0.6$. Technically, this means the results presented here are not entirely self-consistent, as the presented values of [Fe/H] are less than those actually used in our simulation. However, in GCD+ the metallicity only feeds back into the simulation through the cooling and feedback rates, which use the {\em total} metallicity of the particle, regardless of how this metallicity is distributed amongst metal species. As iron only makes up a small portion of a particle's metallicity, the error in a particle's metallicity from this process is $\lesssim5\%$, which is negligible given that the metallicities of gas particles in these simulations varies by two orders of magnitude.


The total disk mass (both gas and stars) is $5\times10^8$ M$_\odot$, and at our fiducial resolution consists of $5\times10^5$ particles, giving a mass resolution of $10^3$ M$_\odot$ per particle, where gas and star particles have the same mass. The stellar disk has a scale height of $100$ pc and a scale length of $860$ pc, while the distribution of gas is set by the criteria of hydrodynamic equilibrium.

\begin{figure}
\begin{center}
\includegraphics[width=\columnwidth]{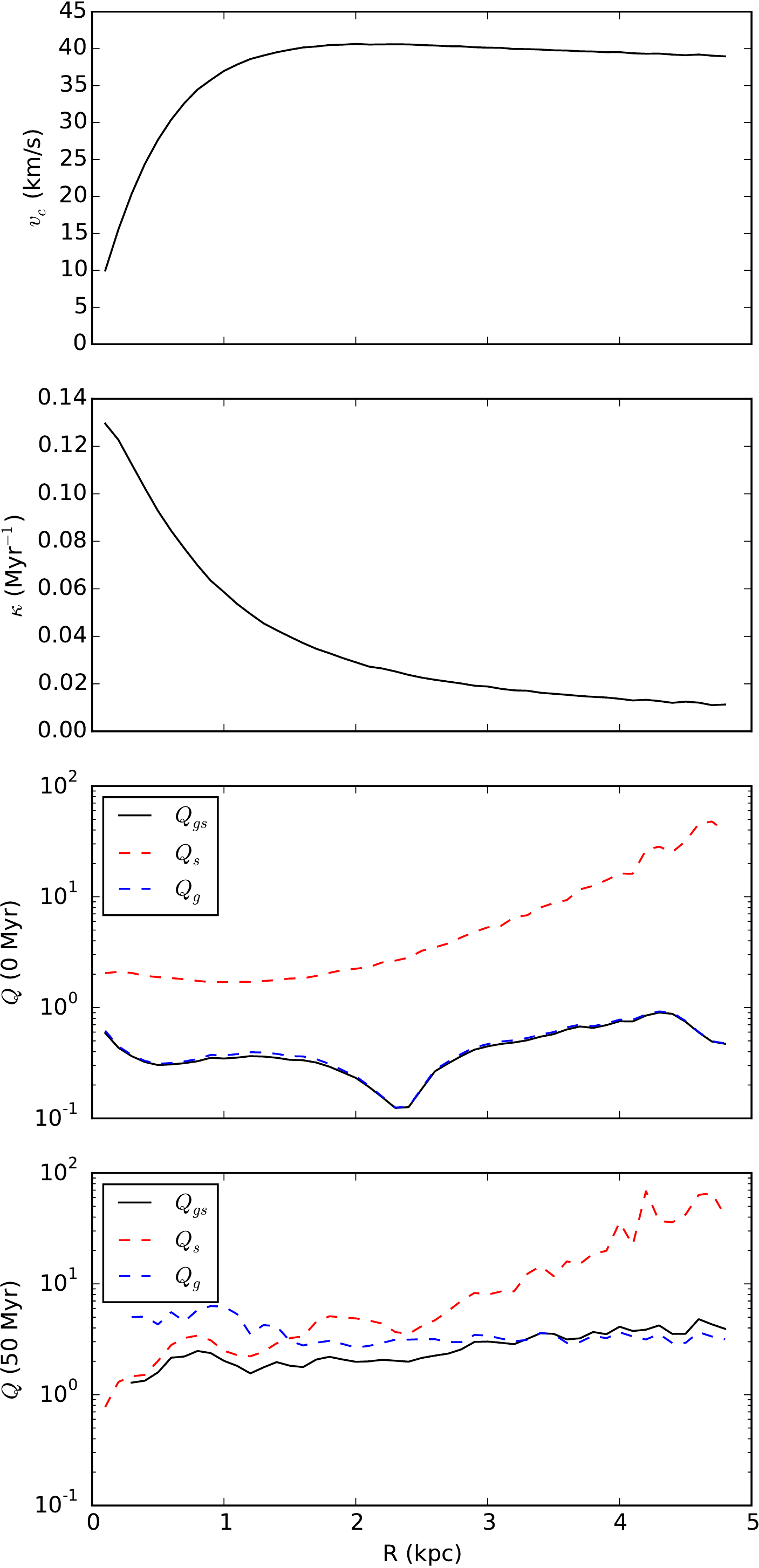}\\
\end{center}
\caption{\label{iniprofs}
From top to bottom: Initial circular velocity ($v_c$) profile; initial epicyclic frequency ($\kappa$) profile; initial $Q$ profile; $Q$ profile at $t=50$ Myr in the ``Shear'' simulation. $Q_\mathrm{s}$ is the $Q$ parameter from the stellar component, $Q_\mathrm{g}$ is the $Q$ parameter from the gaseous component, and $Q_\mathrm{gs}$ is the two-component $Q$ calculated using the method of \citet{2013MNRAS.433.1389R}.
}
\end{figure}

These disks are imbedded within a live dark matter halo of mass $9.5\times10^9$ M$_\odot$, consisting of $9.5\times10^5$ particles. The halo follows a NFW profile \citep{1996ApJ...462..563N} with a concentration of $c=10$. The initial rotation curve, epicyclic frequency, and $Q$ parameters are plotted in Fig.~\ref{iniprofs}. Here we use the two-component $Q_\mathrm{gs}$ of \citet{2013MNRAS.433.1389R}. Although the initial $Q_\mathrm{gs}$ is low, with a minimum of $Q_\mathrm{gs}\sim10^{-1}$, the self-regulating nature of galactic discs causes it to quickly reach a equilibrium of $Q_\mathrm{gs}\approx2$ across the disc.

We also perform a single simulation with $3\times$ our fiducial resolution and a shorter simulation time. We discuss this higher resolution model in appendix~\ref{converge}, where we find that although the strength of feedback is strongly resolution-dependent, metal diffusion is not very sensitive to resolution.

The parameters that vary between the simulations are summarized in Table~\ref{ictable}.

\begin{figure*}
\begin{center}
\includegraphics[width=\textwidth]{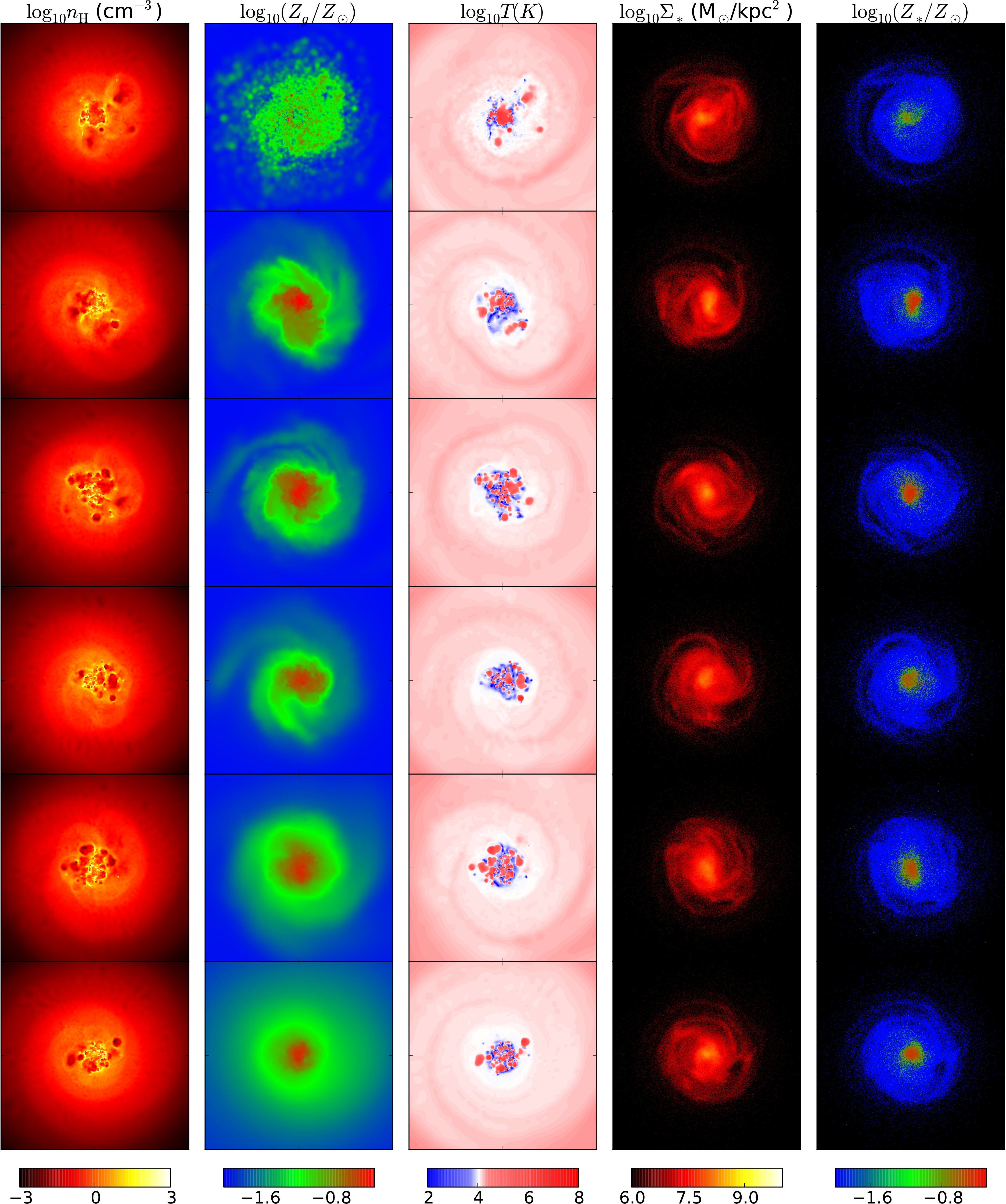}\\
\end{center}
\caption{\label{overview1}
Slices through the $z=0$ kpc plane (i.e. face-on) at $t=750$ Myr, in $5$ kpc x $5$ kpc boxes, for all runs at our fiducial resolution. First column: gas density. Second column: gas metallicity. Third column: gas temperature. Fourth column: star column density. Fifth column: stellar metallicity, mass-averaged along line-of-sight.
}
\end{figure*}

\begin{figure*}
\begin{center}
\includegraphics[width=\textwidth]{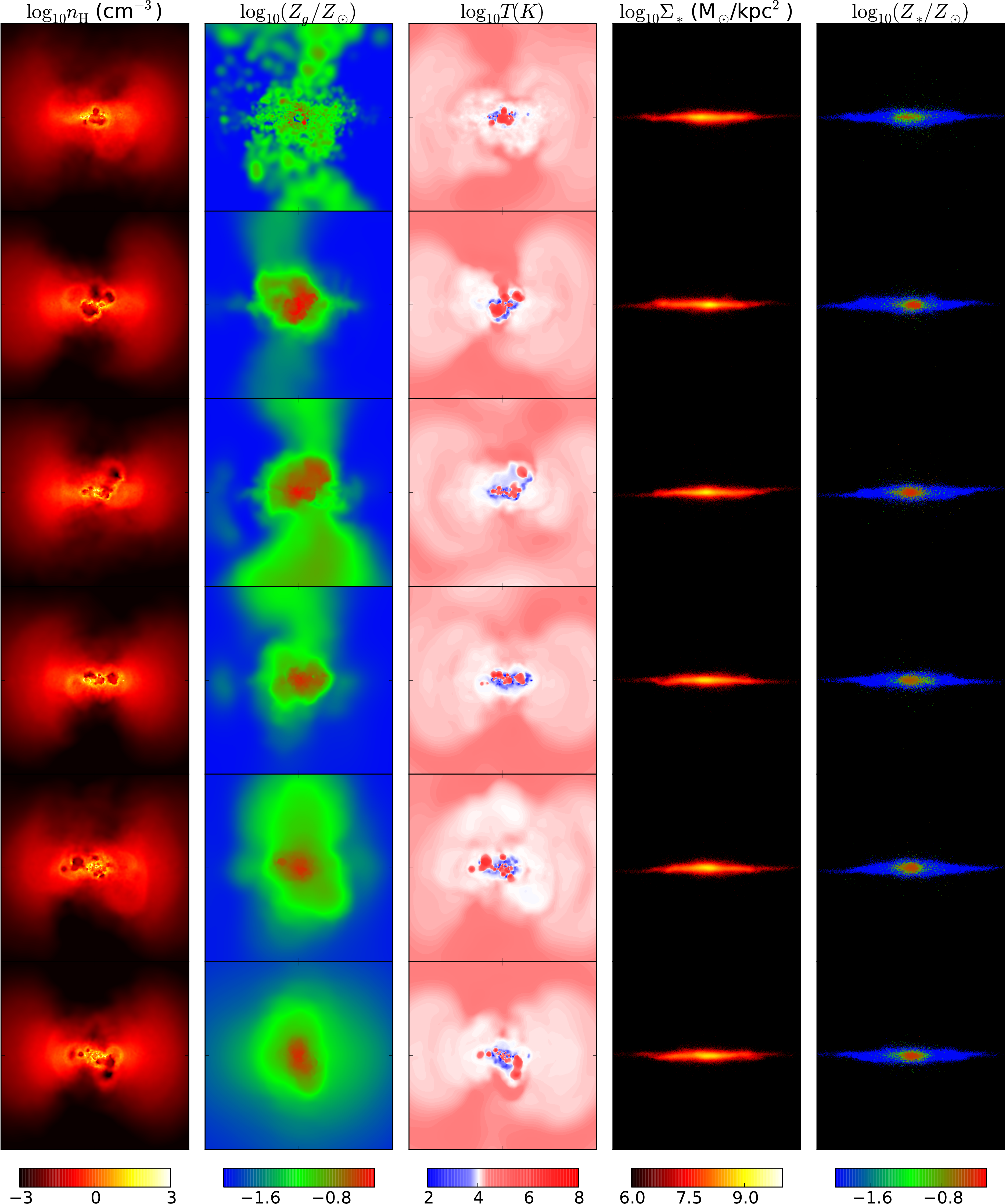}\\
\end{center}
\caption{\label{overview2}
Slices through the $x=0$ kpc plane (i.e. edge-on) at $t=750$ Myr, in $5$ kpc x $5$ kpc boxes, for all runs at our fiducial resolution. First column: gas density. Second column: gas metallicity. Third column: gas temperature. Fourth column: star column density. Fifth column: stellar metallicity, mass-averaged along line-of-sight.
}
\end{figure*}

\begin{figure*}
\begin{center}
\includegraphics[width=\textwidth]{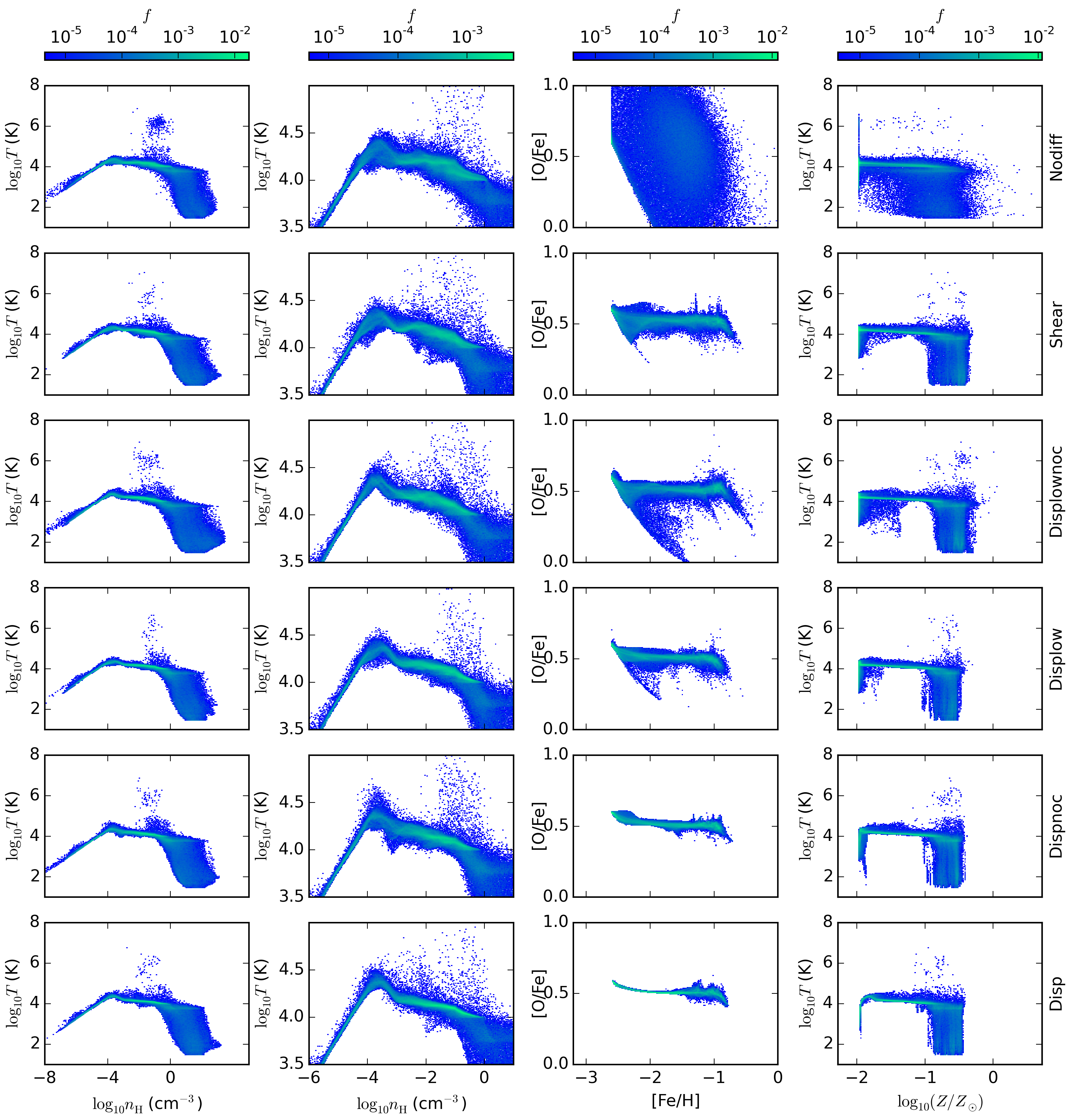}
\end{center}
\caption{\label{lowzphase}
Phase diagrams at $t=750$ Myr. The color bar indicates the fraction of gas mass within each bin. 1st column: $n_\mathrm{H}-T$ phase-plots. 2nd column: a zoom in of the ``equation of state'' region of the 1st column. 3rd column: [O/Fe]-[Fe/H] correlation plot. 4th column: phase plot of metallicity against temperature.
}
\end{figure*}

\section{Results}\label{section_results}

\subsection{General Evolution}

Snapshots at $t=750$ Myr are plotted in Figs.~\ref{overview1} and \ref{overview2}. A metallicity gradient is produced in all models as a result of a greater star formation rate in the central regions of the disk.

These models show a complex thermal structure. Cold clumps and filaments lie within a warm smooth interstellar medium. Hot feedback bubbles are produced within the disk, and this gas can escape, producing a halo of hot gas surrounding the disk. Our initial conditions do not include this hot halo component - it is produced entirely from feedback.

\subsection{Gas properties}\label{gasprop}

The first column of Fig.~\ref{lowzphase} gives $n_\mathrm{H}^{\phantom i}-T$ phase-plots for the simulations at $t=750$ Myr. Most of the gas in all runs follows a tight ``equation of state'' (EoS). This gas is roughly isothermal at $T\sim10^4$ K from $n_\mathrm{H}^{\phantom i}\sim10^{-4}$ cm$^{-3}$ to  $n_\mathrm{H}^{\phantom i}\sim10^{0}$ cm$^{-3}$. At lower densities, the temperature drops, but remains in a tight EoS. This is outflowing gas, where the EoS is regulated by a combination of expansion, and the UV background. At high densities, the gas cools rapidly but some of this gas is also heated rapidly by feedback, producing a broad range of densities and temperatures. Some gas is also heated to high temperature ($T\sim10^{5-7}$ K) producing bubbles, which escape into the halo. These features are common to all models.

\begin{figure*}
\begin{center}
\includegraphics[width=\textwidth]{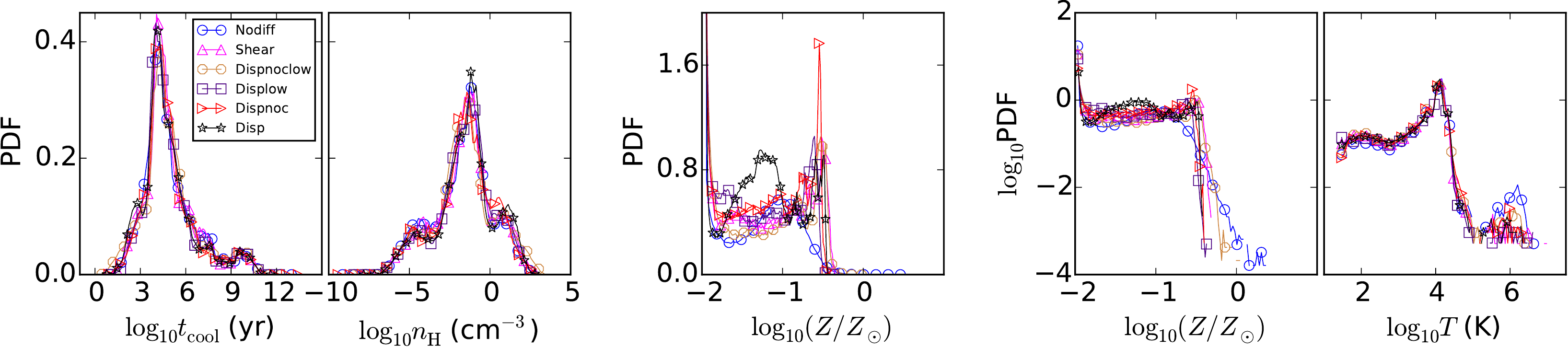}\\
\end{center}
\caption{\label{allhists}
Mass-weighted PDFs of cooling times, gas densities, gas metallicities, and temperatures at $t=750$ Myr for all simulations at the fiducial resolution. The metallicity distribution is plotted both with and without a log scale, to respectively emphasize the high-metallicity and moderate-metallicity ends of the distribution.
}
\end{figure*}

The second column of Fig.~\ref{lowzphase} shows a ``zoomed-in'' view of the EoS region. There is significantly more scatter around the median EoS in Nodiff run than in Disp. We investigated whether this could be due to greater inhomogeneities in the metallicity distribution giving a broader distribution in cooling times when diffusion is weaker or switched off, but Fig.~\ref{allhists} demonstrates that the width of the cooling time distributions are similar in all runs (although having different peaks). Thus it is likely that these are just short-term variations resulting from the current configuration of feedback bubbles, inflows, and outflows.

The third column of Fig.~\ref{lowzphase} shows the [O/Fe]-[Fe/H] phase space for the gas in these models. There is a general negative correlation between [O/Fe] and [Fe/H] in most of the simulations. The correlation is particularly tight in Disp. Such a tight distribution of metallicity has been noted in previous simulations with strong diffusion \citep{2015MNRAS.447.4018G}. The Nodiff models shows no clear correlation at all. Without diffusion, the [O/Fe] ratio of a particle depends only on the feedback events in its immediate vicinity, and so there can be a very large scatter, as different feedback events (e.g. SN Ia vs SN II) occur in different locations in the disk. With strong diffusion, the correlation is tighter, as gas mixes efficiently and the chemical abundances approach the local mean. Indeed, we use the scatter of this correlation to define the strength of diffusion in section~\ref{diffloadsec}.



The fourth column of Fig.~\ref{lowzphase} shows the temperature-metallicity phase space for the gas. The temperature of the bulk of the gas is tightly coupled with the metallicity, and has a slightly lower temperature at higher metallicities, while remaining close to $10^4$ K. Most of the remaining gas (that is, gas significantly above or below $10^4$ K), represents ongoing or recent star-formation. Most of the cold gas exists in cold star-forming regions, while the hot gas is gas that has been heated by feedback - either directly through energy injection, or indirectly through shock-heating. In all models that include diffusion, most of both the cold and the hot gas has a high metallicity, because star-forming regions have the highest metallicity. In the Nodiff model, this gas is spread across all metallicities, because the lack of diffusion allows very low metallicity gas to exist within star-forming regions. In addition to the star-forming cool gas, all models also show that there is some gas at low metallicity and temperature. This is outflowing gas, and its metallicity is low because it has left the disc and is no longer being enriched. Here we can see another correlation with diffusion strength. As diffusion increases in strength, the metallicity of this outflowing gas is reduced. We discuss this further in section~\ref{diffloadsec}.

Fig.~\ref{allhists} shows mass-weighted 1D histograms of the cooling times, density, metallicity, and temperature of all gas in the model at $t=750$ Myr. The distributions for the cooling time, density, and temperature do not change significantly with the diffusion algorithm. When diffusion is present, metal injected by feedback mixes with nearby gas, producing a sharp peak in the PDF at the high metallicity end. More efficient diffusion tends to push this peak to lower metallicities. This trend is produced by enriched gas losing its metals more rapidly. However, this also depends on the recent star formation history, and is not a monotonic trend with diffusion strength. This peak is not present in Nodiff. Instead, there is a high-metallicity tail, consisting of the small number of particles that have been enriched by very many feedback events. The difference between the metal distributions has no clear impact on the distribution of cooling times, presumably because the number of particles in the high-metallicity tail is small.

In general, the gas properties of Shear do not seem to stand out as particularly distinct among the other simulations. The differences between the simulations thus appears to depend only on the strength of diffusion, even when the diffusion coefficient is calculated in different ways. This suggests that diffusion can be accurately calibrated by adjusting the pre-factor, and that it is not necessary to introduce a more complex method for calculating the diffusion coefficient.

\subsection{Spatial Distribution of Metals}

The spatial distribution of metals at $t=750$ Myr is plotted in the 2nd column of Figs.~\ref{overview1} and \ref{overview2}. This distribution is strongly affected by the choice of diffusion algorithm. Without any diffusion, the metal distribution is very clumpy as there is no mechanism to smooth out the metals. Highly metal-enriched particles can therefore move over large distances without losing their metals. Metal-rich gas in the Nodiff model at clearly has a large vertical extent, while Disp is more localized (Fig.~\ref{overview2}).

\begin{figure*}
\begin{center}
\includegraphics[width=\textwidth]{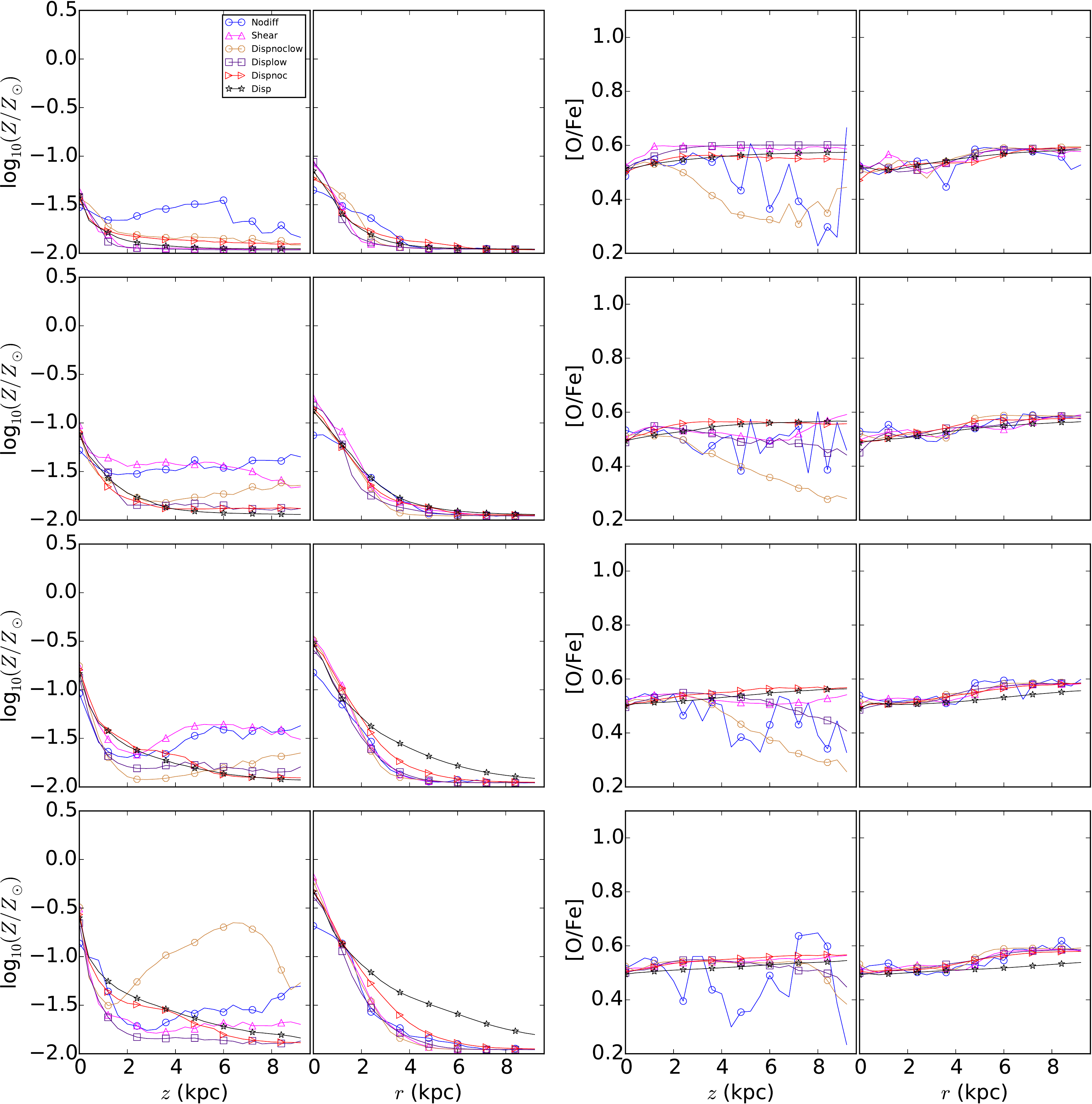}
\end{center}
\caption{\label{metslopes}
Metallicity distributions at $t=250,500,750,$ and $1000$ Myr (from top to bottom).
First column: Metallicity as a function of height above the disk plane. 
Second column: Metallicity as a function of radius. 
Third column: [O/Fe] as a function of height above the disk plane. 
Fourth column: [O/Fe] as a function of radius.
}
\end{figure*}

We have plotted the mean metallicity, and [O/Fe], in vertical and radial bins in Fig.~\ref{metslopes} at intervals of $250$ Myr to examine more clearly the difference between the diffusion algorithms. The radial metallicity bins are annuli that extend $2$ kpc above and below the disk plane, and the vertical metallicity bins are pillboxes of radius $5$ kpc. While many observed dwarf galaxies appear to have flat or near-flat metallicity gradients \citep{2010A&A...516A..85C,2011MNRAS.417.1643K}, all of our simulations show significant slopes. This is likely a result of our feedback being weak, resulting in gas (and hence star formation) that is centrally concentrated. Stronger feedback at higher resolution can produce more distributed star formation (K14). This dependence on feedback means that the simulated spatial distributions of metals can not be directly compared with observations, but only compared with other models with similar feedback strength.

Outside of the central star-forming region, the radial metallicity gradients are similar in most runs. Disp and Dispnoc have shallower slopes at $t=750$ Myr and $t=1000$ Myr, as a result of the strong diffusion transporting metals outwards. This suggests that there is little radial flow of gas, allowing diffusion to dominate.

The vertical gradients show a strong dependence on diffusion. With weak or absent diffusion, we see flat or even positive vertical metallicity gradients, a clear sign of metal-rich outflows. Efficient diffusion strips outflows of their metals before they escape the disk, and so outflows can only be metal-rich (and hence, effective at metal transport) if diffusion is weak \citep{2004MNRAS.352..363M}.

The [O/Fe] profiles follow the inverse of the $Z/Z_\odot$ profiles (i.e. the [O/Fe] slope is positive where the $Z/Z_\odot$ slope is negative and vice versa), due to the negative slope of the correlation noted in section~\ref{gasprop}. The [O/Fe] profiles of Nodiff are Dispnoclow are noisier than the profiles of the simulations with stronger diffusion, due to the greater scatter in the [O/Fe]-[Fe/H] distribution (Fig.~\ref{lowzphase}). 


\begin{figure}
\begin{center}
\includegraphics[width=.98\columnwidth]{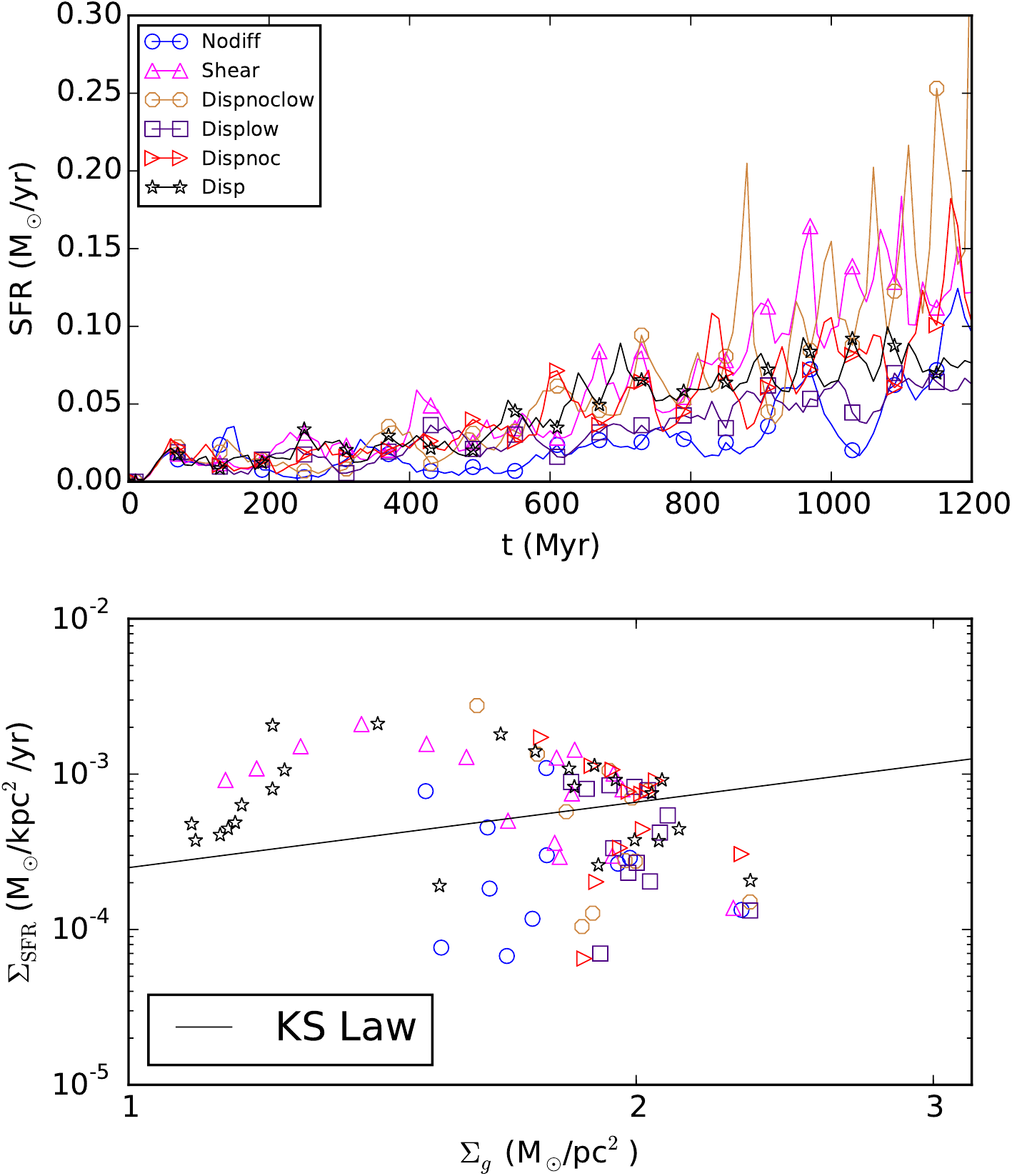}~
\end{center}
\caption{\label{diffhalfrad}
Top: star formation history. Bottom: gas and star formation surface densities with the K-S law. Dots represent samples $100$ Myr apart
}
\end{figure}

We have plotted the star formation rates in Fig.~\ref{diffhalfrad}. All simulations follow similar star formation trends, except for Nodiff which has a significantly lower star formation rate. This is not likely a direct consequence of the diffusion algorithm, and is most likely a result primarily of the chaotic details of the inner region of the galaxy. In other simulations (not presented here) performed with a higher metallicity, we found that the Nodiff simulation produced the {\em highest} star formation rate out of all runs, suggesting that such a variation indeed within the expected range. Fig.~\ref{diffhalfrad} also shows a plot of the galactic mean surface density of star formation as a function of the galactic mean surface density of gas, along with the Kennicutt-Schmidt (KS) law \citep{1959ApJ...129..243S,1998ApJ...498..541K}. The points represent samples $100$ Myr apart within each simulation. Typically, the galaxies move to the left of the plot as gas is consumed or ejected by star formation. There is a large scatter in the star formation rate, but even though it is not likely that the KS law in its standard form extends to these small irregular galaxies, the KS law is still in agreement with many of our star formation rates.

\begin{figure*}
\begin{center}
\includegraphics[width=\textwidth]{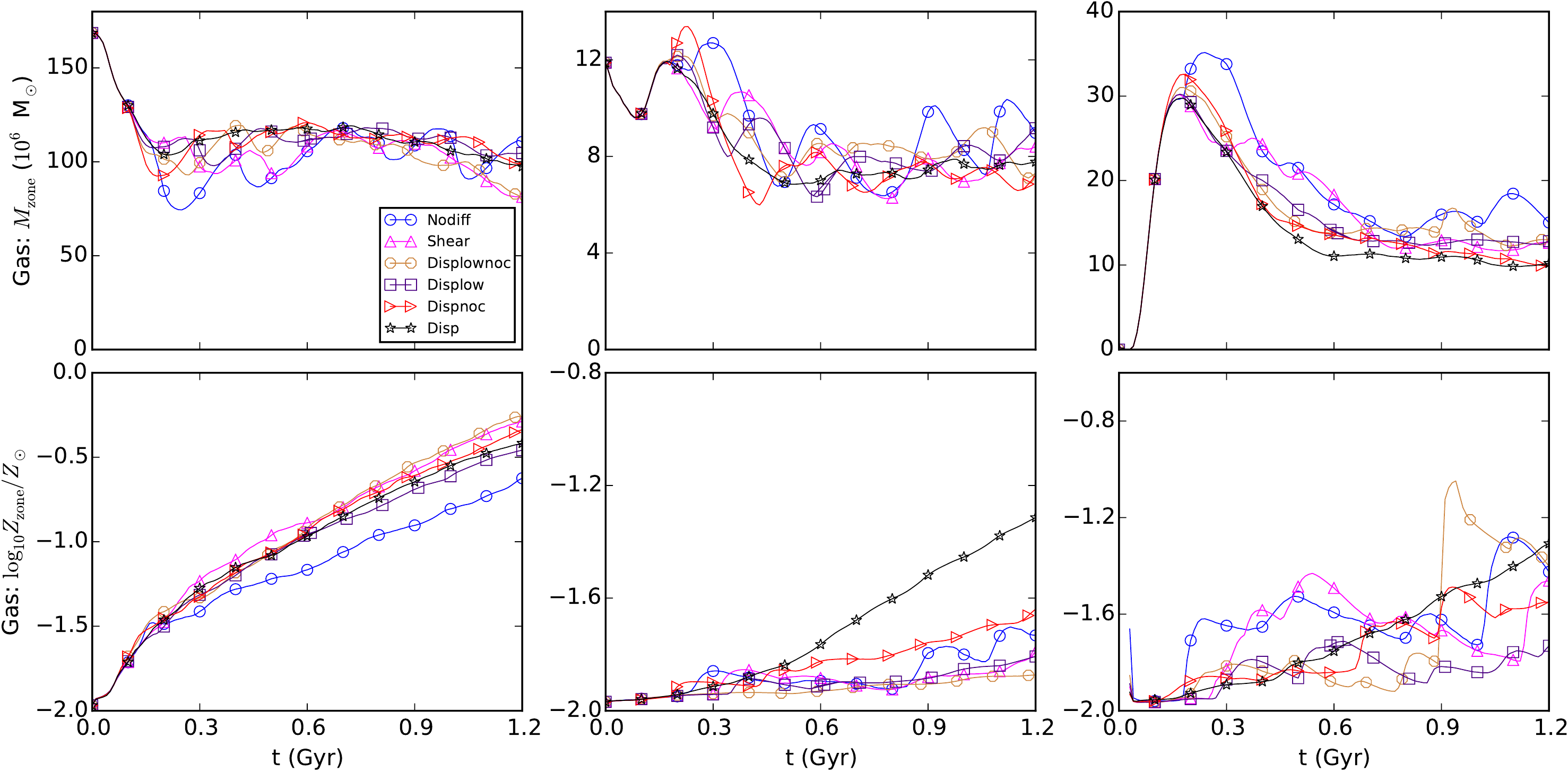}
\end{center}
\caption{\label{metzones}
Top: gas mass in the three zones. Bottom: metallicity in the three zones. Left column: Inner disk zone: $0$~kpc~$\leq R\leq2$~kpc, $|z|\leq2$~kpc. Centre column: Outer disk zone: $4$~kpc$\leq R\leq6$~kpc, $|z|\leq2$~kpc. Right column: Outflow zone: $R\leq6$ kpc, $|z|>2$~kpc.}
\end{figure*}

In Fig.~\ref{metzones} we have plotted the total mass and metallicity of gas as a function of time in three zones - an ``inner disk'' covering the range $0$~kpc~$\leq R\leq2$~kpc and $|z|\leq2$~kpc, an ``outer disk'' over $4$~kpc$\leq R\leq6$~kpc and $|z|\leq2$~kpc, and an ``outflow'' zone where $R\leq6$ kpc and $|z|>2$~kpc. The evolution of the gas mass of each zone does not appear to depend on diffusion strength. Initially, the inner and outer disk zones lose mass to star formation, outflows, and the thickening of the disk, while the outflow zone gains mass. The outflow zone loses mass later as gas falls back onto the disk, or escapes from the galaxy completely, eventually reaching an equilibrium value.

The evolution of the metallicity of each zone reveals more about the effects of diffusion. In the inner disk zone, the metallicity is dominated by star formation, and monotonically increases (with a lower metallicity in Nodiff as a result of its lower star formation rate). In the outer disk zone, metallicities remain low, except in Disp and Dispnoc, where the strong diffusion carries metals outwards. Nodiff also has some increase in metallicity due to its efficient advection of metal-rich gas, but it appears that diffusion dominates over advection in moving metals radially.

In the outflow zone, the metallicity at $t=0$ is undefined as it contains no gas mass. The metallicity of the first gas in the zone varies greatly between the simulations, as it depends on the small number of particles that reach the zone first. After this, the outflow zone gains metallicity at a similar average rate in most simulations, but with very different histories. With strong diffusion (Disp, Dispnoc), the metallicity of the zone gradually increases, as metals ``leak'' upwards through diffusion. With weak or no diffusion, metals are carried upwards by metal-rich outflows, producing a strongly-varying metallicity. The large peaks in the metallicity correspond to small peaks in mass, indicating that it is indeed the mass that is carrying the metals. In Disp, there is no correspondence between peaks in metallicity and peaks in mass, showing that metals are ``leaking'' outwards through diffusion, without any resolved flow of mass. Here, both mechanisms - diffusion and advection - appear to have a similar efficiency at distributing metals, but do so through very different means.

\begin{figure*}
\begin{center}
\includegraphics[width=.95\textwidth]{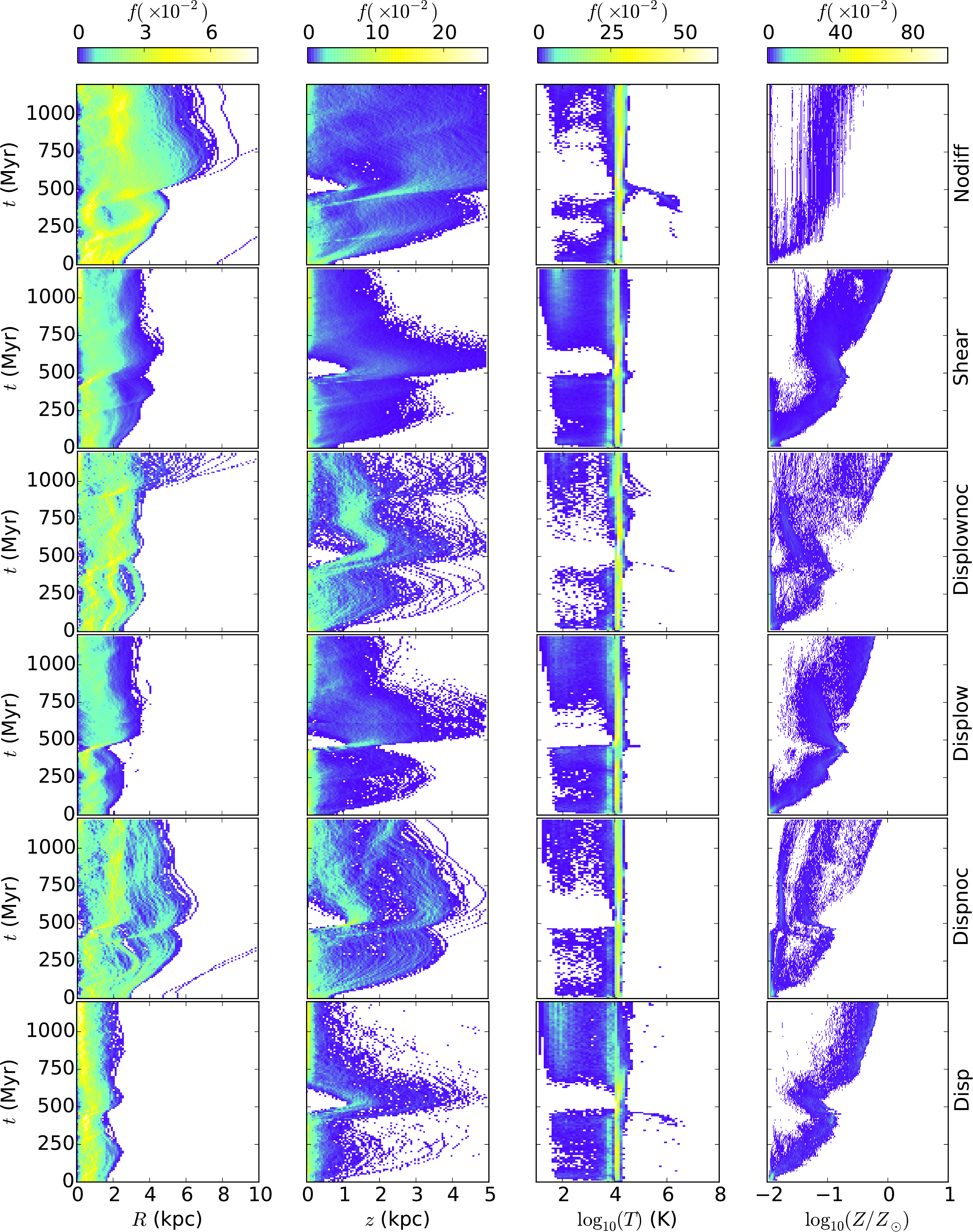}
\end{center}
\caption{\label{radtime1}
Superposition of trajectories of all particles defined as wind particles ($1$ kpc $<|z|<5$ kpc, $|v_z|>10$ km/s) at $500$ Myr. The color bar indicates the fraction of wind particles in each bin at each time.
}
\end{figure*}

\subsection{Wind generation and evolution}

To continue our investigation into the origin of outflowing gas and how this depends on the diffusion method, we identify wind particles at a particular time, and follow these particles forward and backward in time to track the distributions of their metallicity, position, temperature, and vertical velocity. Here we define wind particles as gas particles with $1$ kpc $<|z|<5$ kpc and $|v_z|>10$ km/s. We track wind particles identified at $t=500$ Myr, and plot superpositions of their trajectories through space, temperature, and metallicity over time in Fig.~\ref{radtime1}.

First, we examine the positions of the outflowing particles. The $z$ plots show that most of the outflowing gas traces its origin to the disk, and that in most situations, the outflowing gas appears to mostly come from a single recent outflow event at $t\approx450$ Myr, most of which soon falls back into the disk (although in some cases the outflowing gas appears to to be a composite of a small number of outflow events). The $R$ plots generally show one or two concentrated peaks at this time, showing that most of the gas comes from fairly small regions in the disk. This means that, in general, the outflows are not entraining large quantities of halo gas, but consist primarily of gas ejected directly from the disk, much of which comes from a single concentrated feedback burst.

Next, we examine the temperatures of the outflowing particles. In all runs, most of the particles sit at $T\sim10^4$ K, even before the recent feedback burst. Only a small amount of this gas is below $10^4$ K before the feedback burst (i.e. it is cold, star-forming gas), but this gas is heated by feedback to $10^4$ K. Additionally, a small quantity of gas is heated to $T\sim10^6$ K in some runs, but this quickly cools to $T\sim10^4$ K. Together, this shows most of the outflowing gas was not cold star-forming gas, but was part of the warm ISM that was blown out by a feedback bubble. That is, outflows are effective at entraining nearby gas, although only within a small region near the feedback burst.

Finally, we examine the metallicities of the outflowing particles, where the effect of diffusion becomes the most clear. We see a ``base-line'' of the tracked gas at the initial metallicity ($\log_{10} Z/Z_\odot=-2$), that persists throughout the simulation in the absence of diffusion. With diffusion, this base-line disappears (more quickly with stronger diffusion), as the enriched gas mixes throughout the disk. This is only the case for the outflowing gas we track: if we similarly track all galactic gas instead of certain selected particles, we would find that a significant quantity of  gas remains at the initial metallicity, showing that not all the galaxy's gas receives enrichment, even with strong diffusion.

Without diffusion, individual particles gain metallicity and retain it without mixing, giving no clear pattern for the evolution of the gas as a whole, and producing metallicities that can only increase. With diffusion, we see the gas is enriched, and reaches a peak in metallicity before being ejected -- this is metal injection from the feedback burst. Following this, the enriched ejected gas loses metallicity because it is no longer receiving metals from feedback, and the metals can diffuse into the CGM and ISM of the galaxy. The strongest diffusion models tend to cause a more rapid drop in the outflowing gas metallicity. Most of the gas then falls back into the disk, where it continues the process of being enriched.

\subsection{Stellar properties}\label{starprop}

We have shown that different strengths of diffusion produce different spatial distributions of metals in the gas component, but this spatial distribution depends strongly on our star formation and feedback algorithm, and our hydrodynamic scheme. This introduces a degeneracy when comparing simulations to observations. However, a comparison of the properties of stars among our simulations should be less sensitive to these factors, if we only examine stars that are formed during the simulation, and neglect stars that are present in the initial conditions.

\begin{figure}
\begin{center}
\includegraphics[width=\columnwidth]{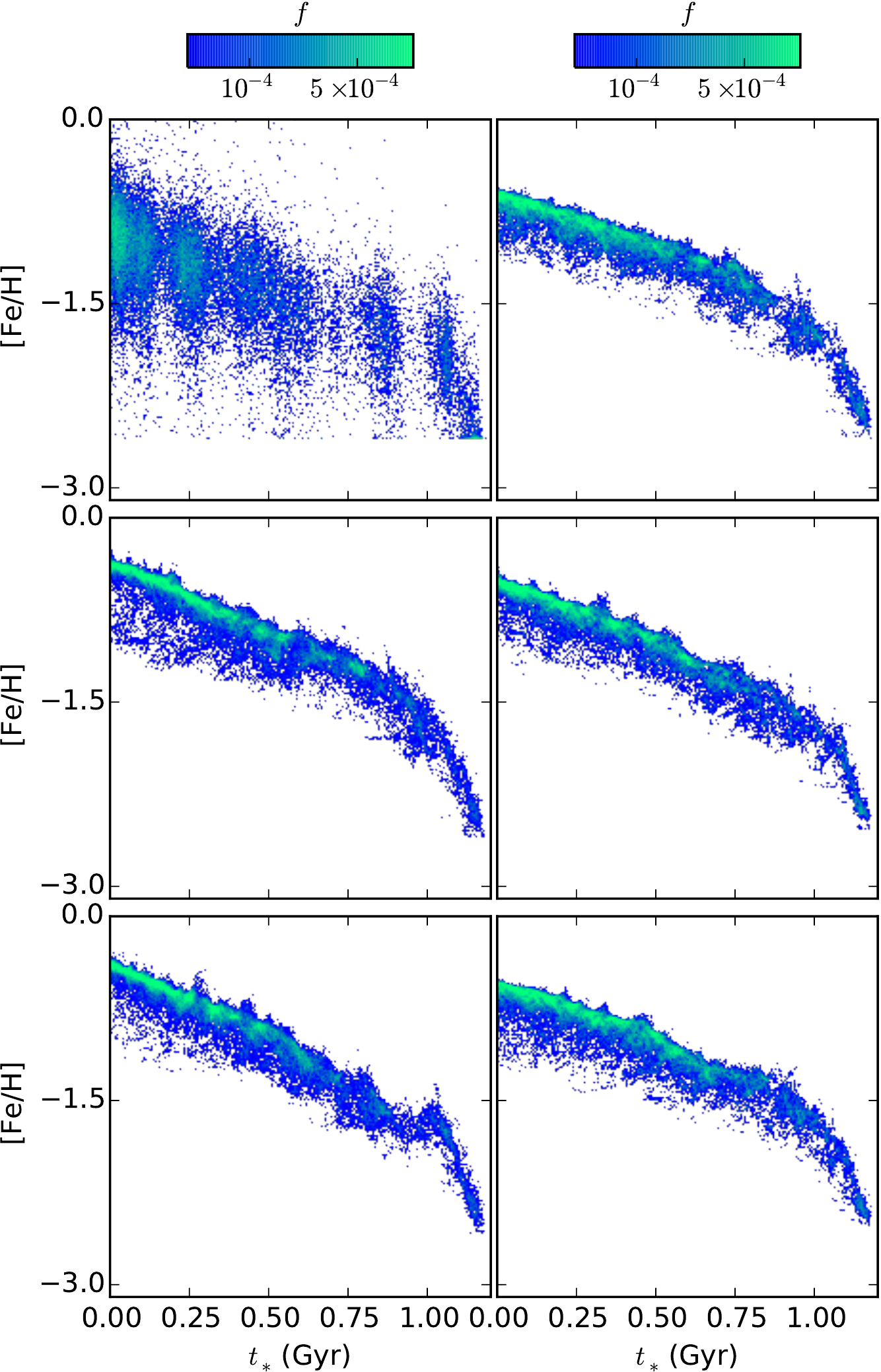}
\end{center}
\caption{\label{metage}
Age-metallicity relation for stars formed by $t=1200$ Myr.. Top row: Nodiff, Displow. Middle Row: Shear, Dispnoc. Bottom row: Displownoc, Disp. The color bar indicates the fraction of stars in each bin. 
}
\end{figure}

We plot the age-metallicity relation for stars formed by $t=1$ Gyr in Fig.~\ref{metage}. Despite the dramatic variations in the distribution of metals in the gas, the age-metallicity relations of stars have remarkably similar slopes and variances, with the exception of Nodiff, which has an extremely broad distribution. In these simulations, stars are formed primarily in a small central region of the galaxy, where even the weaker diffusion algorithms can produce efficient mixing.

\begin{figure}
\begin{center}
\includegraphics[width=\columnwidth]{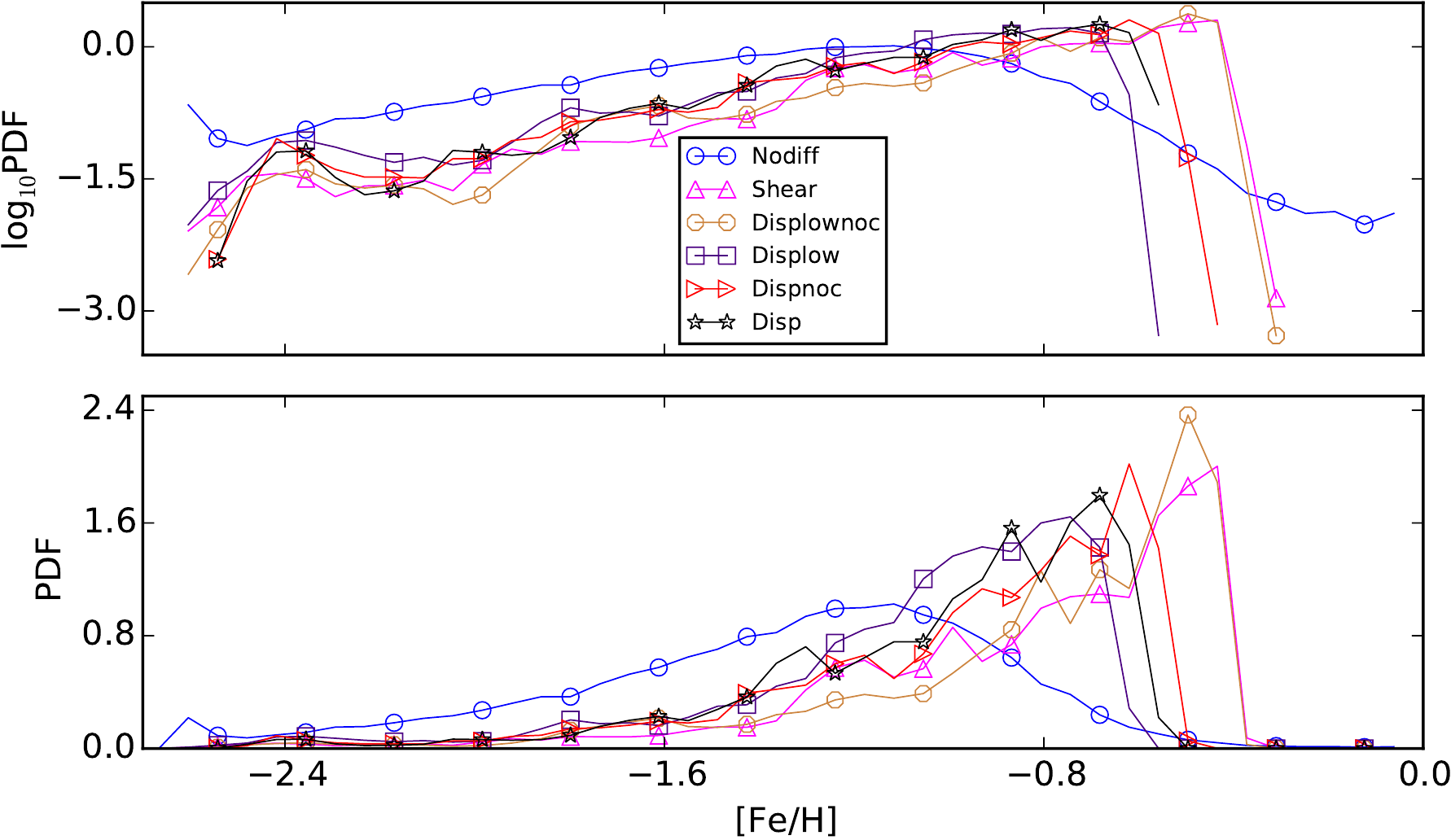}
\end{center}
\caption{\label{starmdf}
Metallicity distribution functions for stars formed by $t=1200$ Myr. The metallicity distribution is plotted both with (top) and without (bottom) a log scale, to respectively emphasize the high-metallicity and moderate-metallicity ends of the distribution
}
\end{figure}

The metallicity distribution function of stars formed by $t=1$ Gyr is plotted in Fig.~\ref{starmdf}. Most stars are formed in the central high-metallicity region, giving a high metallicity peak to the MDF. As with the gas MDFs (Fig.~\ref{allhists}, discussed in section~\ref{gasprop}), there is a weak trend for the high-Z end of the MDF to be truncated at lower metallicities with stronger diffusion. Nodiff is a strong outlier, with stars forming at much lower metallicites than in any of the simulations with diffusion. The suppression of the formation of low-metallicity stars by diffusion has already been noted in the literature \citep{2012MNRAS.425..969P}, but here we can make the further conclusion that this effect does not seem to depend on the strength of diffusion - even the weakest diffusion produces an MDF that is very different to that of Nodiff. 

In general, and in contrast to the gas, the stellar population does not have a strong dependency on the diffusion strength, provided that some diffusion is present. Feedback is likely a more significant effect, as we comment on in appendix~\ref{converge}.

\subsection{Time-averaged effects}\label{diffloadsec}

The episodic nature of star formation in dwarf galaxies produces strong but short-lived variations in the properties of galactic gas and outflows. This makes it difficult to determine which effects are due to the difference in diffusion strength, and which are due to short-term variations. By averaging properties over the first $1.2$ Gyr of the simulation, we can more clearly discern the differences caused by changing the diffusion strength.

We use the scatter of the [O/Fe]-[Fe/H] relation to parameterize the strength of diffusion - stronger diffusion results in a smaller scatter. For each simulation, we divide the gas population at $t=750$ Myr into $20$ bins according to their [Fe/H] value, and calculate the standard deviation of [O/Fe] in each bin. The mean of these standard deviations is the ``[O/Fe]-[Fe/H] scatter'' that we use here. We plot this against the time-averaged outflow mass-loading, outflow metal mass-loading, and star formation rate. The outflow mass-loading and metal mass-loading are calculated by tracking all gas particles that pass from $|z|\le2$ kpc and $R<10$ kpc to $z>2$ kpc or $R>10$ kpc between output dumps before $t=1.2$ Gyr. We note that the mass-loading does depend on this threshold height, and that the choice of $2$ kpc is somewhat arbitrary.

\begin{figure}
\begin{center}
\includegraphics[width=\columnwidth]{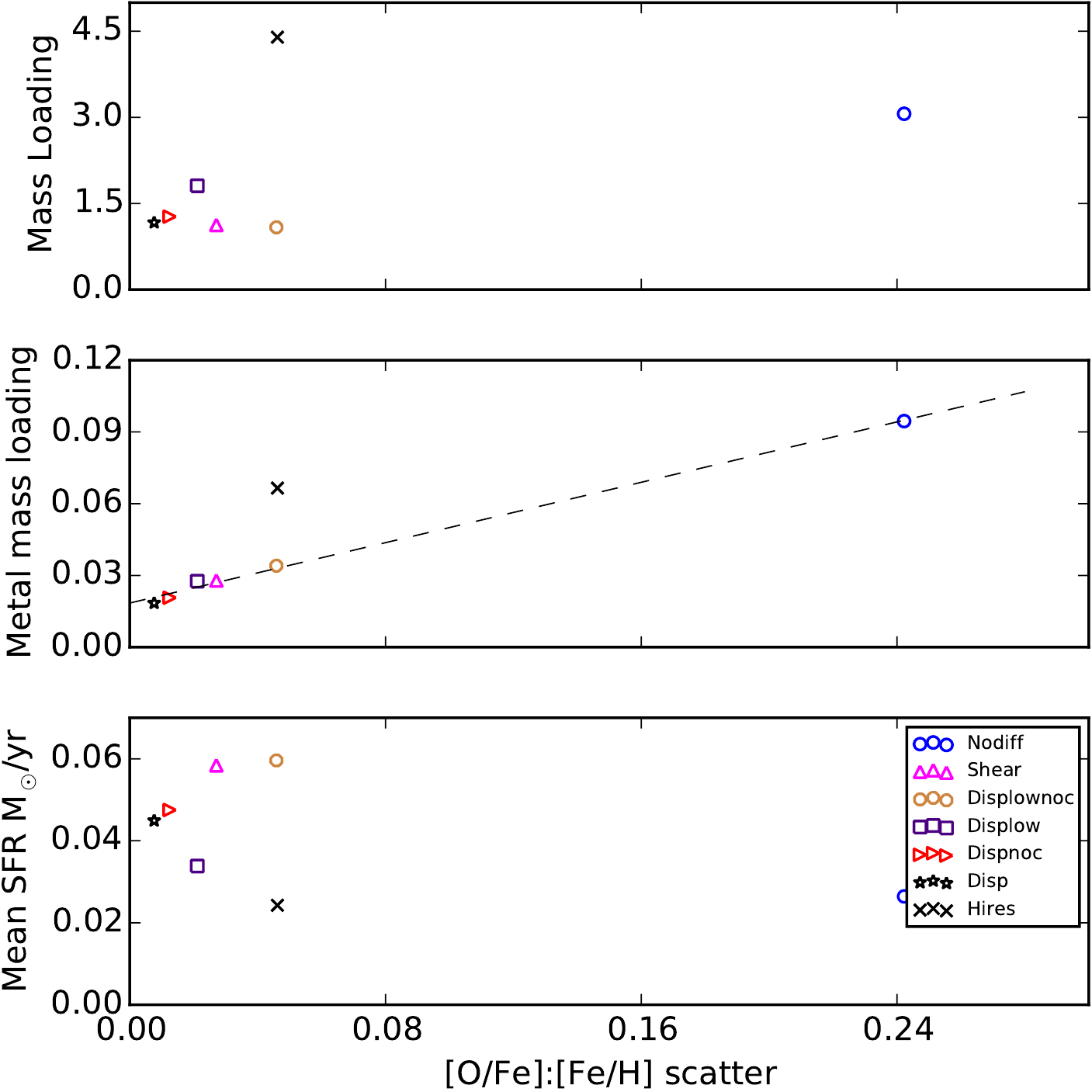}
\end{center}
\caption{\label{diffload}
Mass-loading (outflow mass divided by total star formation), metal mass-loading (outflow metal mass divided by total metal production), and star-formation rate, time-averaged over the first $t=1.2$ Gyr of the simulation, except for Hires which had a simulation time of $t=770$ Myr.
}
\end{figure}

The total mass and total metal mass that passes through the threshold over the entire simulation are divided by the total star formation mass and the total mass of metals produced respectively, to produce the mean mass-loading and metal mass-loading across the simulation. These results, along with the time-averaged SFR, are plotted in Fig.~\ref{diffload}. This figure also includes results for the Hires simulation, but we exclude this from our analysis here, and discuss its results in appendix~\ref{converge}.

The mass-loading appears almost constant across all runs at our fiducial resolution that include diffusion. Displow has a somewhat higher mass-loading, while Nodiff has a very large mass-loading. However, the metal mass-loading shows a remarkably tight correlation with [O/Fe]-[Fe/H] scatter, fitting a linear slope of $y=0.316x+0.018$. Here we can clearly see the main effect of diffusion on outflows: diffusion reduces the metal mass-loading of enriched outflows, because the diffusion removes metals from metal-rich gas and transports it to metal-poor gas. While it has already been shown that strong diffusion can strip outflows of their metals and reduce their effectiveness at transporting metals \citep{2004MNRAS.352..363M}, it is still surprising to find such an incredibly tight correlation between diffusion strength and metal mass-loading.

It is not clear whether there is such a correlation with the star formation rate. Although Disp, Dispnoc, Shear, and Displownoc appear to follow a linear trend, Nodiff and Displow have significantly lower mean star formation rates. These two simulations are also the ones that had a larger than average mass-loading. This leads us to the conclusion that there may be a weak correlation between star formation rate and [O/Fe]-[Fe/H] scatter, but that this can be overwhelmed if the gas happens to be arranged such that feedback can produce particularly efficient outflows.

We note that, as in section~\ref{gasprop}, Shear is not an outlier among the other simulations. This reinforces the conclusion that modifying the method for calculating the diffusion coefficient does not produce significantly different results to simply setting a different value for the pre-factor.

\section{Discussion of hydrodynamic model}\label{hydrodiscuss}

In SPH, only feedback events with enough energy to propel a significant number of SPH particle masses can produce a resolved advecting outflow. Our particle mass is $1000$ M$_\odot$, and so each feedback event must blow out many thousands of solar masses to be resolved by at least one smoothing kernel. It could be useful to make use of self-consistent sub-grid turbulent models \citep[e.g.][]{2009MNRAS.398..548B} to better capture these unresolved processes, and their contribution to metal diffusion. Although grid-based Eulerian simulations include implicit mixing, this numerical diffusion can be stronger than should be expected for laminar flows, while potentially weaker than would be expected from sub-grid turbulence.

Sub-grid turbulence models have already been implemented in ISM and IGM models \citep[e.g.][]{2009MNRAS.398..548B,2010MNRAS.405.1634S}, but resimulating our models in such a drastically different code is beyond the scope of this paper. Instead, as a ``sanity check'', we produce a simple estimate for the time-scale of one potential source of turbulence, and of the diffusion time-scale based on observational data.

In our first model, we explore unresolved instabilities in galactic bubbles as a source of turbulence. We model feedback as producing a bubble of size $R$ of hot gas with a dense cool bubble wall. This situation is particularly sensitive to the RT instability.

At early times, the RT instability grows at a rate of approximately
\begin{equation}h(t)\approx\alpha A g t^2,\end{equation}where $h(t)$ is the magnitude of the instability, $g$ is the external gravity field, $t$ is time, $\alpha$ is a constant of order unity, and $A$ is the Atwood number, defined as
\begin{equation}A=\frac{\rho_w-\rho_b}{\rho_w+\rho_b},\end{equation}where $\rho_w$ and $\rho_b$ are the densities of the wall and bubble gas respectively. For this simple order-of-magnitude estimate, we can set $A=1$,  because $\rho_w\gg\rho_b$. We also set $\alpha=1$. Thus the bubble growth rate depends on the gravity field $g$. We set $g$ equal to the gravity from our NFW halo near the galaxy centre. When the distance from the centre of an NFW potential is smaller than the scale-length of the potential, the gravitational acceleration approaches a constant. In our models, this is $g\sim3\times10^{-8}$ cm s$^{-2}$. We can define the instability as being extremely significant when its magnitude $h$ approaches the radius of the feedback bubble, $R_\mathrm{fb}$. Thus we can derive an approximate RT instability time-scale $\tau_\mathrm{RT}^{\phantom i}$,
\begin{equation}\tau_\mathrm{RT}^{\phantom i} \sim 3 \mathrm{~Myr} \left(\frac{R_\mathrm{fb}}{100 \mathrm{~pc}}\right)^{1/2}\end{equation}
The instability has a time-scale that is less than or comparable to the dynamical times of the gas, suggesting that the RT instability could provide a significant source of turbulence in our simulations. 

We can look to observations to estimate the turbulent diffusion time-scale without the requirement of understanding its source. The 21-cm data of \citet{2008MNRAS.387L..18R} show that the non-thermal contribution to the line-width $v_\mathrm{nt}$ of H$\textsc{i}$ gas scales as approximately

\begin{equation}
\frac{v_\mathrm{nt}}{1\mathrm{~km~s}^{-1}}=\beta \left(\frac{l}{1\mathrm{~pc}}\right)^{0.35},
\end{equation}where $l$ is a length-scale, and $\beta\approx1$ for gas with a pressure of $2000$ cm$^{-3}$~K. Defining a time-scale for turbulent diffusion as $\tau_T^{\phantom i}=l/v_\mathrm{nt}$, we therefore find that

\begin{equation}
\tau_T^{\phantom i}\sim (20\mathrm{~Myr}) \left(\frac{l}{100\mathrm{~pc}}\right)^{0.65}.
\end{equation}
This is comparable to the time-scales of resolved hydrodynamical processes, confirming that unresolved turbulence is likely to be significant.

\section{Conclusions}\label{section_conc}

We have performed smoothed particle hydrodynamics simulations of isolated dwarf galaxies with five different diffusion methods (resulting in five different diffusion strengths) and one simulation without any diffusion. By comparing the properties and evolution of the gas and stars, we can make several conclusions:

\begin{enumerate}
\item The [O/Fe]-[Fe/H] distribution of the gas has a much smaller scatter as diffusion increases in strength.
\item Diffusion strips high-metallicity gas of its enrichment before it can escape the disk, producing a remarkably tight correlation between diffusion strength (as measured by the scatter in the [O/Fe]-[Fe/H] relation) and the metal mass-loading of outflows.
\item The simulation with Shear diffusion is not an outlier amongst the Disp simulations, suggesting that the gas properties depend only on the strength of diffusion, even if the diffusion coefficient is calculated in different methods. This implies that adjusting the pre-factor of the diffusion coefficient could have an equivalent effect.
\item This is a weak trend for stronger diffusion to truncate both the gas and stellar MDF at lower metallicities.
\item The stellar age-metallicity relation and the stellar MDF do not strongly depend on diffusion strength (outside the high-Z end), provided some diffusion is present.
\item Strong diffusion allows metals to ``leak'' out of the galactic disk without any explicitly-resolved mass flow, while with weak diffusion, outflows retain their metals and can be very effective at transporting metals, producing flat or even positive vertical metallicity gradients in the gas.
\end{enumerate}

As the diffusion strength has a critical role in determining which processes are dominant in a galaxy, future work must ensure that the diffusion model is selected carefully. Perhaps particular care should be taken when using Eulerian codes, where numerical diffusion can be strong. Sub-grid turbulence models may provide a useful method for precisely determining the strength of diffusion in a less phenomenological fashion.

\section*{Acknowledgements}
This research was supported by the Canada Research Chair program and NSERC. Simulations were run on the Calcul-Qu\'{e}bec/Compute-Canada supercomputers {\em Colosse} and {\em Guillimin}. We thank Alessandro Romeo for useful discussions that contributed to this paper.\\

\bibliographystyle{apj}
\bibliography{metsec}

\appendix

\section{Convergence Check}\label{converge}

\begin{figure*}
\begin{center}
\includegraphics[width=.98\textwidth]{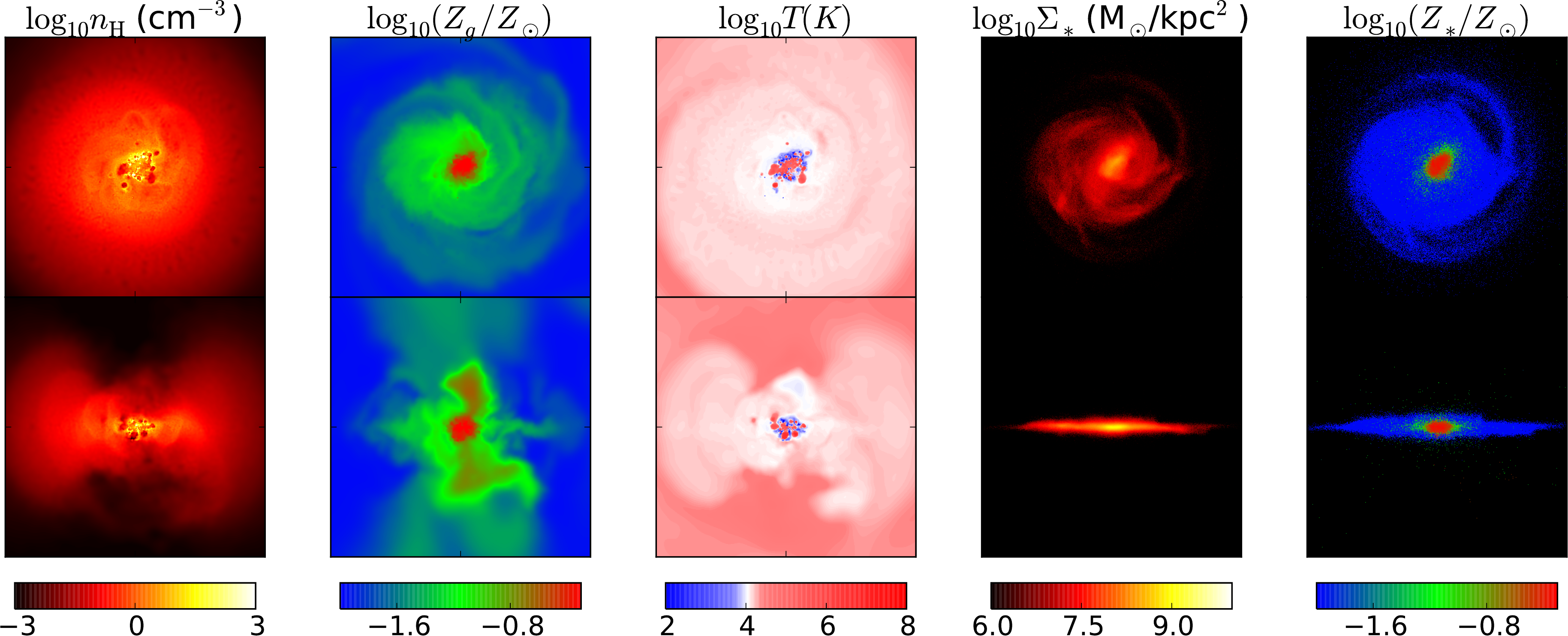}
\end{center}
\caption{\label{hislice1}
Summary of the Hires simulation. Top: Slices through the $x=0$ kpc plane (i.e. edge-on). Bottom: Slices through the $z=0$ kpc plane (i.e. face-on). All slices are $5$ kpc x $5$ kpc boxes, taken at $t=750$ Myr. Left column: gas density. Second column: gas metallicity. Third column: gas temperature. Fourth column: star column density. Fifth column: stellar metallicity, mass-averaged along line-of-sight.
}
\end{figure*}

\begin{figure*}
\begin{center}
\includegraphics[width=.98\textwidth]{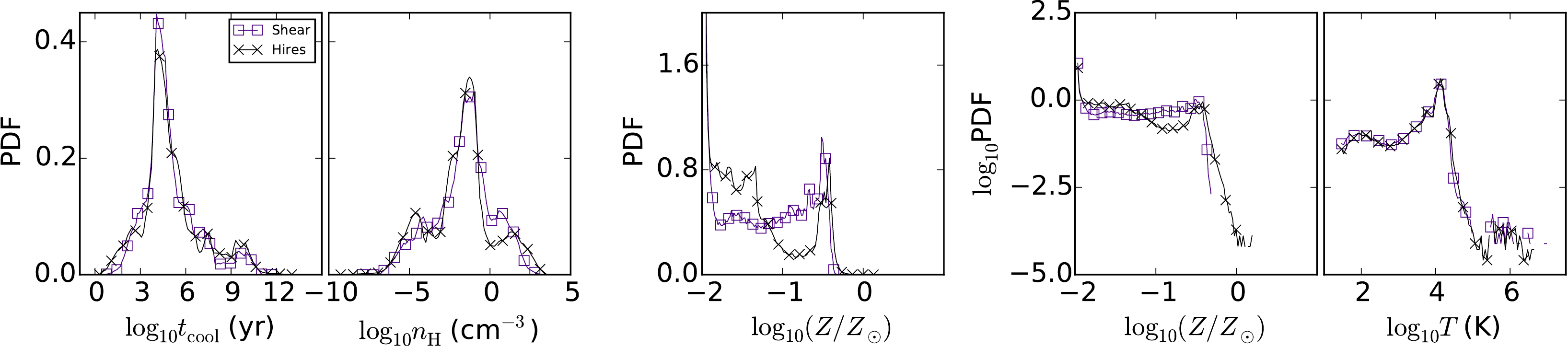}
\end{center}
\caption{\label{hihists}
Mass-weighted PDFs of cooling times, gas densities, gas metallicities, and temperatures at $t=750$ Myr for Hires and Shear.
}
\end{figure*}

\begin{figure}
\begin{center}
\includegraphics[width=.98\columnwidth]{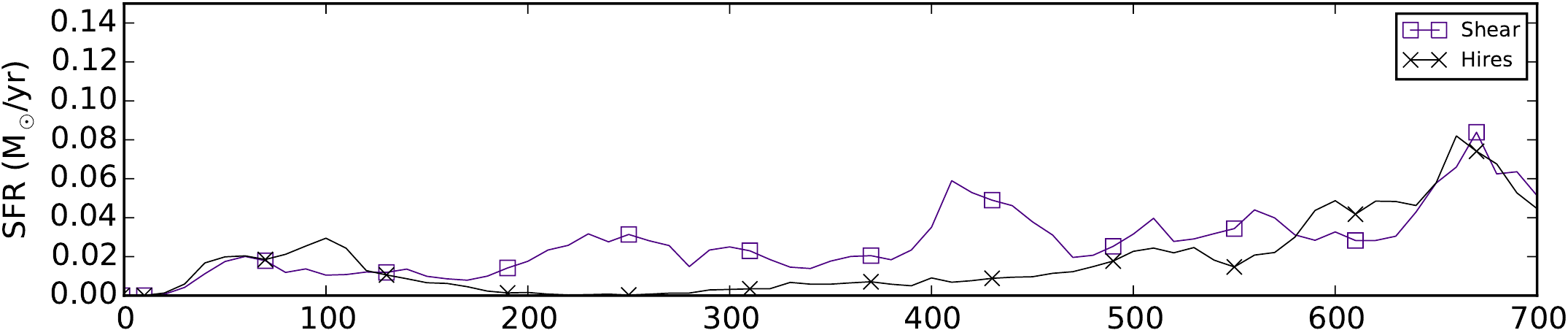}
\end{center}
\caption{\label{hisfr}
Star formation rates for Shear and Hires.
}
\end{figure}

We performed one additional run at three times the resolution of our standard runs, using the same diffusion model as the ``Shear'' simulation. Slices taken at $t=750$ Myr are plotted in Fig.~\ref{hislice1}. Both the cold gas and hot gas bubbles are more centrally concentrated than in Shear (4th row of Figs.~\ref{overview1} and \ref{overview2}). The metallicity of the outflows also appears to be higher in Hires than in other simulations.

We have plotted mass-weighted PDFs of the of the gas properties of Hires and Shear at $t=750$ Myr in Fig.~\ref{hihists}. The temperature, cooling time, and density PDFs are similar, but the metallicity distribution reaches a lower maximum metallicity in Hires. This is primarily a result of the lower star formation rate (Fig.~\ref{hisfr}), and can be explained by feedback being more efficient at higher resolutions.

Fig.~\ref{diffload} shows that the metal mass-loading and mass-loading factors of Hires are much larger than the general trend, confirming that feedback is producing more powerful outflows. These plots also show that the [O/Fe]-[Fe/H] scatter of Hires is only a little higher than in Shear. Thus, the increased metal mass-loading is not a result of weaker diffusion, but primarily as a result of the increased efficiency of feedback. With a lower gas particle mass, the production of feedback bubbles is better resolved, and the pressure of feedback particles more efficiently couples with the kinetic energy of outflows. Gas is ejected more rapidly, and has less time to lose metals through diffusion, producing more enriched outflows.

The weak dependence of the [O/Fe]-[Fe/H] scatter on resolution shows that our conclusions about diffusion remain valid, and that our diffusion algorithm is not strongly resolution-dependent. However, the actual mass-loading and metal mass-loading of the outflow depends strongly on the strength of feedback, which we have found to be strongly dependent on resolution.

\section{Idealized diffusion}\label{diffcheck}

\begin{figure*}
\begin{center}
\includegraphics[width=.98\textwidth]{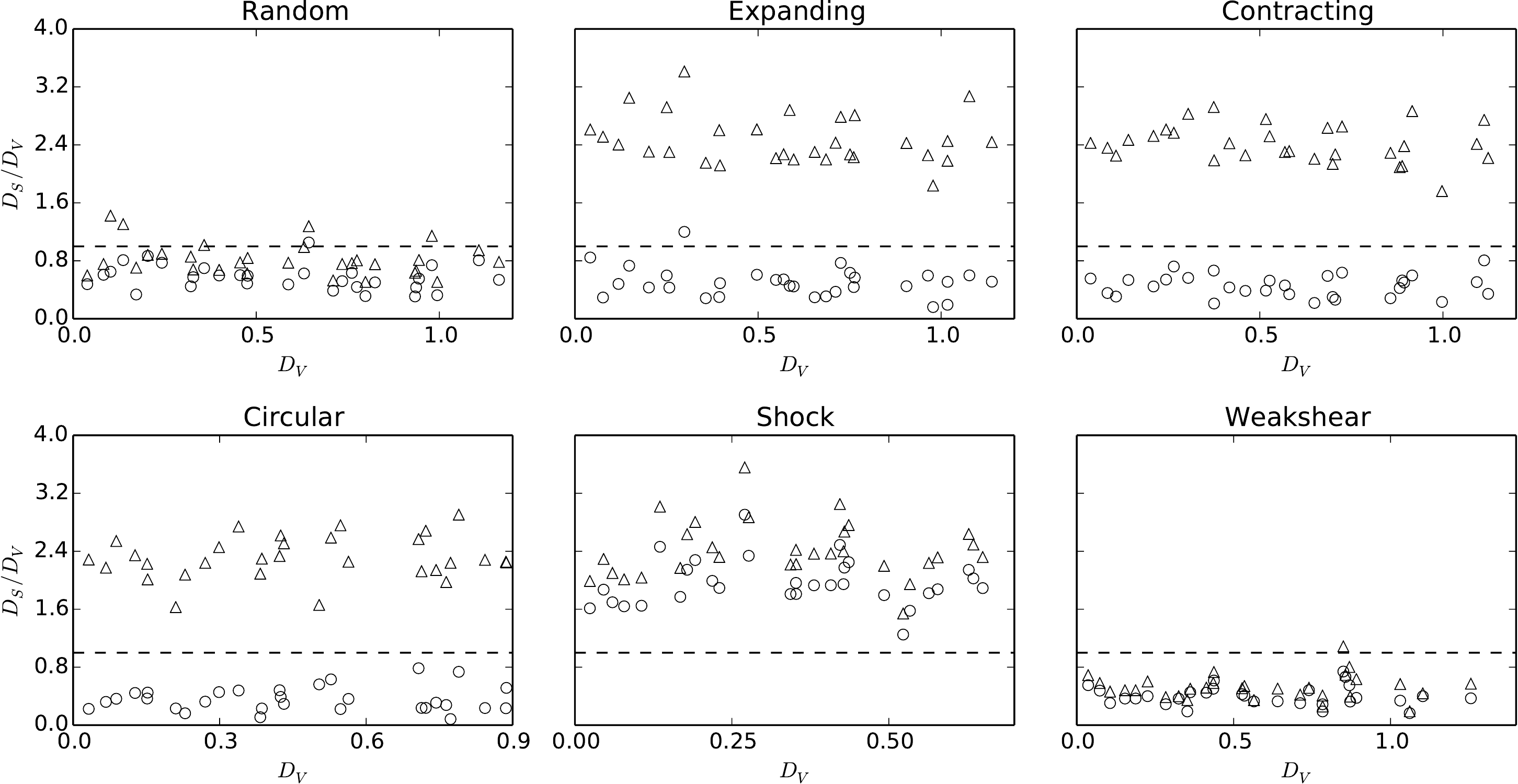}
\end{center}
\caption{\label{diffcomp}
Ratio of $D_S$, the diffusion coefficient calculated using the ``shear'' method, to $D_V$, the diffusion coefficient calculated using the ``velocity dispersion'' method, as a function of $D_V$, for a variety of idealized distributions of particle velocities. Circles represent $D_S$ calculated using the standard Shear method (as in S10), where the velocity scale is the symmetric trace-free shear tensor. Triangles represent $D_S$ calculated in the same fashion, except that the shear tensor is not symmetrized and does not have its trace removed. For particle positions $\mathbf{r}$, and a velocity scale $f_v$, the velocities $\mathbf{v}$ for each distribution are as follows. Random: $\mathbf{v}=f_v(\alpha,\beta,\gamma)$, where $\alpha,\beta,\gamma$ are three independent random numbers in the range $(-1,1)$. Expanding: $\mathbf{v}=f_v\mathbf{r}$. Contracting: $\mathbf{v}=-f_v\mathbf{r}$. Circular: $\mathbf{v}=f_v(y,-x,0)$. Shock: $\mathbf{v}=f_v(-x,0,0)$. Weakshear: $\mathbf{v}=f_v(|x|,|y|,|z|)$
}
\end{figure*}

To quantify the inherent differences between the diffusion algorithms, we calculated the diffusion coefficient in a variety of idealized scenarios using three methods:

\begin{enumerate}
\item Shear with the standard trace-free symmetric shear tensor.\\
\item Shear but without removing the trace or symmetrising the shear tensor, i.e. $S_{ab,i}=\tilde{S}_{ab,i}$.\\
\item Dispnoc (i.e. based on the velocity dispersion, and excluding the sound-speed term).\\
\end{enumerate}
This is to show the effects of removing the trace and symmetrising the shear tensor compared to simply using the velocity dispersion. We generated $64$ particles with positions randomly distributed within a cube that extends from $x=y=z=-1$ to $x=y=z=1$, and a series of different velocity distributions, detailed in Fig.~\ref{diffcomp}. Here, we have set $C_{DS}=C_{DV}=1$ for simplicity, and set the smoothing length to be equal to the distance to the most distant particle. We produce a number of different random realizations of particle positions, varying the velocity scale $f_v$ from $f_v=0.1$ to $f_v=3.0$, and plot each realization of particle positions as a point. 

In most situations, we find that $D_S\approx(1/2)D_V$. As the Shear method sums up velocity differences, while the Disp method sums up the square of velocity differences, it makes sense that the Shear method gives lower values. We note that the diffusion coefficient does not go to zero for the trace-free symmetric shear tensor in the case of purely circular, contracting, expanding motion. This is due to inhomogeneities in the density distribution producing a momentum field that is not purely circular, contracting, or expanding. Similarly, the ``Weakshear'' velocity distribution has deliberately been designed to minimize the velocity gradients, but the density inhomogeneities permit significant diffusion to remain. With a uniform grid of particles, we find that $D_S=0$ in these situations.

Expansion along one axis - which we call the ``Shock'' velocity distribution - is the one case where $D_S$ is significantly greater than $D_V$. Removing the trace from the shear tensor is designed to remove the contributions of spherically symmetric expansion or contraction, but has very little effect in the case of expansion or contraction that is primarily along one axis. This is not an uncommon situation in simulated galaxies, and will occur when particles pass through a shock with a width much greater than the smoothing length. Although it is difficult to argue which diffusion method is more ``physical'' without more comparisons to experiment and observation, it seems that in the ``shock'' case, the Shear method is producing larger diffusion coefficients than was intended, and that perhaps an additional switch is required to alleviate this.

\end{document}